\documentclass[twocolumn]{autart}    

\usepackage{savesym}
\savesymbol{AND}

\usepackage[pdftex]{graphicx}
\usepackage{amsmath}
\usepackage{amssymb}
\usepackage{txfonts}
\usepackage{bm} 
\usepackage{subcaption}
\usepackage{algorithmic,epstopdf}
\usepackage{algorithm}
\usepackage{multirow}
\usepackage{todonotes}
\usepackage{color}
\usepackage{url}
\usepackage{soul}

\newcommand{\LOne}{\mathcal{L}_1[0,\infty]}
\newcommand{\InPro}[2]{\left\langle{#1},{#2}\right\rangle}

\DeclareMathOperator*{\argmin}{argmin}

\newtheorem{example}{Example}
\begin{document}

\begin{frontmatter}

\title{On the Coordinate Change to the First-Order Spline Kernel for Regularized Impulse Response Estimation \thanksref{footnoteinfo}} 

\thanks[footnoteinfo]{This paper was not presented at any IFAC
meeting. Corresponding author Y.~Fujimoto. Tel. +81-93-695-3545.}

\author[Kyoto]{Yusuke Fujimoto}\ead{y-fujimoto@kitakyu-u.ac.jp}    
\ and \author[Shenzhen]{Tianshi Chen}\ead{tchen@cuhk.edu.cn}

\address[Kyoto]{Faculty of Environmental Engineering, The University of Kitakyushu, Wakamatsu-ku, Kitakyushu, 808-0135, Japan}  %
\address[Shenzhen]{
School of Science and Engineering and Shenzhen Research Institute of Big Data, The Chinese University of HongKong,
Shenzhen, 518172, China}

\begin{keyword}                           
Identification methods,
kernel-based regularization methods,
impulse response estimation, kernels.             
\end{keyword}                             

\begin{abstract}
The so-called tuned-correlated kernel (sometimes also called the first-order stable spline kernel) is one of the most widely used kernels for the regularized impulse response estimation. This kernel can be derived by applying an exponential decay function as a coordinate change to the first-order spline kernel.
This paper focuses on this coordinate change and derives new kernels by
investigating other coordinate changes induced by stable and strictly proper transfer functions.
It is shown that the corresponding kernels inherit properties from these coordinate changes and the first-order spline kernel. In particular, they have the maximum entropy property and moreover, the inverse of their Gram matrices has sparse structure. In addition, the spectral analysis of some special kernels are provided. Finally, a numerical example is given to show the efficacy of the proposed kernel.
\end{abstract}

\end{frontmatter}

\section{INTRODUCTION} \label{sec:intro}
One of the main difficulties in system identification is to balance
the data fit and the model complexity \cite{Ljung:1999}. Recently, a
new method to handle this issue is proposed by Pillonetto and
De Nicolao, especially for the impulse response estimation
of linear time-invariant systems \cite{Pillonetto:2010}. Their main idea comes from
the regression over the Reproducing Kernel Hilbert Space
(RKHS) \cite{Aronszajn:1950,Scholkopf:2001} in the machine learning field. These spacecs are related to bivariate
functions that are called kernels and this class of methods is often referred to as the kernel-based regularization methods.
In contrast with the classical Prediction Error Methods (PEMs), a property of such methods is that it
is possible to design through the kernel a model structure that contains a wide class of impulse responses.
More specifically, recall that the classical PEMs first determines the model structure,
and then tune its parameters according to the observed data.
In this case, the set of all possible impulse responses is a finite dimensional
manifold. On the other hand, the kernel-based regularization method, with a carefully designed kernel, searches the impulse response within a possibly infinite dimensional RKHS and thus has the potential to model complex systems.

%

One of the main issues for the kernel-based regularization method is how to design a suitable kernel.
While various kernels have been proposed (e.g., \cite{Prando:2015,Chen:2015-4,Chen:2018}), three
most widely used kernels are the so-called Stable Spline
kernel (SS) \cite{Pillonetto:2010}, the Tuned-Correlated kernel (sometimes also called the first-order stable spline kernel) \cite{Chen:2012}, and the Diagonal-Correlated
(DC) kernel \cite{Chen:2012}. These three kernels have simple structures
and favorable properties, and their effectiveness are shown
in various works, e.g., \cite{Chen:2012,Pillonetto:2014,Bottegal:2015,Prando:2016}.


Interestingly, these three kernels share some common properties
\cite{Chen:2018,Chen:2018-2}.
For example, they can be derived by applying an exponential decay function as a coordinate change to different kinds of spline kernels \cite{Wahba:1990} (cf. Section ~\ref{subsec:probset} for details).
Moreover,
they also inherit some properties from the corresponding spline kernel \cite{Chen:2016}, such as the maximum entropy (MaxEnt) property and the spectral analysis.

Based on the above observations, the following questions then arise naturally:
\begin{itemize}

\item Instead of the exponential decay coordinate change, can we design kernels with other type of coordinate change suitable for system identification?

\item What is the corresponding a priori knowledge embedded in such kernels?

\end{itemize}

%

In this paper, we aim to address the above questions. In particular,  we will focus on the kernels derived by applying
the impulse response of a stable and strictly proper transfer function $G(s)$ as the coordinate
change function to the first-order spline kernel, where $s$ is the complex frequency for the Laplace transform.
Then it is obvious to see that the exponential decay function $e^{-\alpha t}$ is a special case of the proposed kernels with $G(s)=\frac{1}{s+\alpha}$.
Besides, in our preliminary work \cite{Fujimoto:2017}, we considered the case
where the coordinate change is given by $t^ne^{-\alpha t}$, which corresponds to
$G(s)=\frac{1}{(s+\alpha)^n}$. Here, we will consider more general cases and moreover, we will show that such coordinate change embeds a priori knowledge from $G(s)$ on the regularized impulse response, or equivalently,
the corresponding RKHS inherits properties from $G(s)$. For instance, the proposed kernels are
always stable, and the estimated impulse response has the
same convergence rate as the coordinate change function. The relative degree of the impulse response
is determined by the coordinate change function. Morevoer, we
also show that the proposed kernels have the Maximum Entropy
property and give the spectral analysis for some special cases based on the corresponding ones for the first-order spline kernel.

The remaining part of this paper is organized as follows.
Sec.~\ref{sec:probset} recaps the kernel-based regularization methods, and {states} the problem
considered in this paper.
{Sec.~\mbox{\ref{sec:positiveDefinite}} first shows the positive definiteness and stability of the proposed kernel.} Then
Sec.~\ref{sec:ZeroCross} shows properties of the proposed kernels
related to zero-crossing.
Sec.~\ref{sec:maxEnt} discusses the Maximum Entropy property of the proposed kernel,
and Sec.~\ref{sec:SAofMPSK} gives spectral analysis for some
special cases.
Sec.~\ref{sec:sim} shows a numerical example to
demonstrate the effectiveness of coordinate changes.
{Finally Sec.~\mbox{\ref{sec:conclusion}} concludes this paper}.

\noindent
[Notations]
{Sets of
nonnegative real numbers and natural numbers are
denoted by
$\mathbb{R}_{0+}$ and $\mathbb{N}$, respectively. }
The {$n$-dimensional} identity matrix is denoted by $I_n$.
The inverse and the transposition of a matrix $A$ are
denoted by $A^{-1}$ and $A^{\top}$, respectively.
The determinant of a square matrix $A$ is denoted by $\det(A)$.
$\|A \|_{\rm FRO}$ denotes the Frobenius norm of matrix $A$.
The $(i,j)$ element of a matrix $A$
is denoted by $\{A\}_{i,j}$.
When $a$ is a vector,
$\{a\}_i$ denotes the $i$\,th element
of $a$.
The Lebesgue integral of $f(x)$ over $\mathcal{X}$ is denoted by
$\int_{\mathcal{X}}f(x) dx$,
and the integral with the measure $\mu$ is denoted by
$\int_{\mathcal{X}}f(x) d\mu(x)$.
In particular,
the Lebesgue integral of $f(x)$
over $[a,b)$ is denoted
by
$\int_a^b f(x)dx$.
$\LOne$ shows the
set of absolute integrable functions
over $[0,\infty)$,
i.e.,
$\LOne=\{f\mid \int_0^\infty |f(x)|dx<\infty\}.$
The set $\{a_1,\ldots, a_n\}$ is denoted by
$\{a_k\}_{k=1}^n.$
{The expected value and variance of
random variables are denoted by $\mathbb{E}$ and
$\mathbb{V}$}, respectively.
{The limit $\lim_{t\to +0}f(t) $ denotes the right-sided limit at zero. }
Throughout the paper, $s$ denotes the
complex frequency for the Laplace transform,
and $e$ denotes the
Napier's constant.


\section{PROBLEM SETTING} \label{sec:probset}

%
%
%
%
%
%

\subsection{Kernel-based regularization methods}
We first recap the
kernel-based regularized method for continuous-time
systems.
This paper focuses on single-input-single-output, bounded-input-bounded-output, 
stable, linear
time invariant and causal systems described by
\begin{align}
y(t) = \int_0^{t} u(t-\tau) g(\tau) d\tau +w(t), \label{eq:system}
\end{align}
where $t \in \mathbb{R}_{0+}$ is the time index, {$u(t)\in \mathbb{R}, y(t)\in \mathbb{R},$ and $w(t)\in \mathbb{R}$ }are the input, the measured output,
and the {measurement} noise at time $t$, respectively,
$g(t):\mathbb{R}_{0+}\to \mathbb{R}$ is the impulse response of the system, and
$\int_0^{t} u(t-\tau) g(\tau) d\tau $ is the convolution
of the input and the impulse response,
$w(t)$ is {independently and identically Gaussian distributed with mean 0 and variance $\sigma^2$}.
The identification problem in this paper
is to {estimate} $g(t)$
from
the measured output $\left\{y(t_k) \right\}_{k=1}^N$ and the input
$u(t)$ {over the interval $[0,t_N]$}, {where $t_1, t_2, \ldots, t_N$ are the sampling time instants}.

To this end, we use the kernel-based regularization method where the estimated impulse response $\hat{g}(t)$ is given by
\begin{align}
\hat{g}=\argmin_{g \in \mathcal{H}} \sum_{k=1}^N\left(y(t_k)-\int_0^{t_k}u(t_k-\tau)g(\tau)d\tau \right)^2+
\gamma \|g\|_{\mathcal{H}}^2. \label{eq:representer_identification}
\end{align}
Here, $\mathcal{H}$ is a Hilbert space of
functions $g:\mathbb{R}_{0+}\to\mathbb{R}$, and
$\| \cdot \|_{\mathcal{H}}$ is the norm endowed to $\mathcal{H}$, and
$\gamma>0$ is a regularization parameter. Clearly, a good estimate of the impulse response depends on a good choice of $\mathcal{H}$. In the sequel, we assume that $\mathcal{H}$ is a
Reproducing Kernel Hilbert Space (RKHS).

The definitions of RKHS and the reproducing kernel are as follows.
Let $\mathcal{X}$ be a nonempty set, and
consider the Hilbert space of functions $f:\mathcal{X}\to \mathbb{R}$
denoted by $\mathcal{H}$. Then $\mathcal{H}$ is a RKHS if 
\begin{align}
\forall x \in \mathcal X, \exists C_x: |g(x)|\leq C_x \|g\|_{\mathcal{H}},\quad \forall g \in \mathcal{H}.
\end{align}
Further let $\InPro{\cdot}{\cdot}$ be the inner product endowed to $\mathcal{H}$. Then
a symmetric bivariate function $K:\mathcal{X}\times \mathcal{X}\to \mathbb{R}$ is the reproducing kernel of $\mathcal{H}$ if it satisfies
\begin{align}
f(x) = \InPro{f}{K(x,\cdot)}, \quad \forall f \in \mathcal{H},
\end{align}
where $K(x,\cdot)$ indicates the single-variable function defined by
setting the first argument of $K$ to $x$. Reproducing kernels are also called kernels for short.
It is well-known that the kernel $K$ exists if the Hilbert space $\mathcal H$ is RKHS.

With the above definitions, the optimal solution of (\ref{eq:representer_identification}) has explicit expression.
Let $K: \mathbb{R}_{0+} \times \mathbb{R}_{0+} \to \mathbb{R}$
be the kernel of $\mathcal{H}$ in (\ref{eq:representer_identification}).
Also let $O \in \mathbb{R}^{N \times N}$ be a
matrix which is defined as
\begin{align}
\{O\}_{i,j}=\int_0^{t_i} u(t_i-\tau_1)  \int_0^{t_j} u(t_j-\tau_2) K(\tau_1,\tau_2) d\tau_1 d\tau_2.
\end{align}
Let
$y = [y(t_1), \ldots, y(t_N)]^{\top} \in \mathbb{R}^N$
and $c \in \mathbb{R}^N$ be
\begin{align}
c = (O+\gamma I_N )^{-1}y. \label{eq:c}
\end{align}
Then, the optimal solution of (\ref{eq:representer_identification})
is given by
\begin{align}
\hat{g}(t)
= \sum_{i=1}^N \{c\}_i K_i^u(t),  \label{Eq:ghat}
\end{align}
where
$K_i^u(t)$ is a function of $t$
defined by
\begin{align}
K_i^u(t) = \int_0^{t_i} u(t_i-\tau)
K(\tau, t) d\tau. \label{eq:ConvKernelInput}
\end{align}
See e.g., \cite{Pillonetto:2014} for more {details}.


\subsection{Problem statement} \label{subsec:probset}

The Stable Spline (SS) kernel, the Tuned-Correlated kernel (sometimes also called the first-order stable spline kernel), and the DC kernel can all be derived by applying an exponential decay function as a coordinate change to different kinds of spline kernels. To make this point clear, we let $W(\cdot,\cdot)$ be a kernel function and $X(t)$ be a coordinate change function. Then the aforementioned three kernels can all be
put into the following form
\begin{align}
K(t_1,t_2) = W(X(t_1),X(t_2)),\quad t_1,t_2\in \mathbb{R}_{0+}.
\end{align} Moreover, the coordinate change functions $X(t)$ are all  $e^{-\alpha t}: \mathbb{R}_{0+}\to [0,1]$ for these three kernels, while the kernel $W$ is the
second order spline kernel for the SS kernel, the first order spline kernel for the TC kernel, and a generalized first order spline kernel for the DC kernel, cf. \cite{Chen:2016-2}.
In particular, the TC kernel,
\begin{align}
K_{{\rm TC}}(t_1,t_2) = \beta \min(e^{-\alpha t_1}, e^{-\alpha t_2}), \quad t_1,t_2\in \mathbb{R}_{0+}
\label{eq:defTC}
\end{align}
can be derived by applying $e^{-\alpha t}: \mathbb{R}_{0+}\to [0,1]$
as the coordinate change to the first-order spline kernel
\begin{align}\label{eq:spline-od1}
K_{S}(\tau_1,\tau_2) = \beta \min(\tau_1,\tau_2), \quad \tau_1,\tau_2\in [0,\ 1],
\end{align}
where $\beta>0$ and $\alpha>0$ are hyperparameters of {the} kernel.
We consider more general coordinate changes in this paper.

\begin{prob}
Let $G_0(s)$ be a stable and strictly proper transfer function and $g_0(t):\mathbb{R}_{0+} \to \mathbb{R}$
be the impulse response of $G_0(s)$.
Hereafter, {we consider properties of the kernel
given by the
first-order spline kernel with
$|g_0(t)|$ as the coordinate change function, i.e.},
\begin{align}
K_{G_0}(\tau_1,\tau_2) = \min (|g_0(\tau_1)|,|g_0(\tau_2)|). \label{eq:G}
\end{align}
or equivalently the properties of the RKHS associated with $K_{G_0}$ that is denoted by $\mathcal{H}_{G_0}$ below.
\end{prob}



%

\section{{POSITIVE DEFINITENESS AND STABILITY}} \label{sec:positiveDefinite}

We first recall some definitions.

A kernel $K : \mathcal{X} \times
\mathcal{X} \to \mathbb{R}$ is said to be \textit{positive definite} if 
the Gram matrix of $K$ defined as 
\begin{align}
\begin{bmatrix}
K(x_1,x_1) &K(x_1,x_2) &\cdots &K(x_1,x_m) \\
K(x_2,x_1) &  &&K(x_2,x_m) \\
\vdots &&\ddots&\vdots \\
K(x_m,x_1) &K(x_m,x_2)&\cdots& K(x_m,x_m)
\end{bmatrix} \in \mathbb{R}^{m\times m}, \label{eq:defGram}
\end{align}
is positive semidefinite
for any $[x_1, \ldots, x_m]^{\top} \in \mathcal{X}^m$ and for any $m \in \mathbb{N}$. 
The Moore-Aronszajin theorem states that if $K$ is positive definite, then there exists a unique RKHS whose reproducing kernel is $K$ \cite{Aronszajn:1950}.

A kernel $K:\mathbb{R}_{0+} \times \mathbb{R}_{0+}\to \mathbb{R}$ is said to be stable if $\mathcal{H}$, the RKHS associated with $K$,
satisfies $\mathcal{H} \subset \LOne$.

Then we have the following result\footnote{All proofs of propositions are deferred to the Appendix.}.

\begin{thm}
The kernel (\ref{eq:G}) is positive definite and moreover, stable, i.e., the corresponding RKHS $\mathcal{H}_{G_0}$ is a subspace of $\LOne$.
\label{thm:pd_stability} \end{thm}

Theorem~\ref{thm:pd_stability} shows that the kernel (\ref{eq:G}) is a positive semidefinite kernel and moreover,
for any $g\in\mathcal H_{G_0}$, $g\in\LOne$.

\section{ZERO-CROSSING RELATED PROPERTIES} \label{sec:ZeroCross}

%
%
%
%

The following proposition
shows that if $g_0(t)$ has a zero-crossing,
then any $g \in \mathcal{H}_{G_0}$ inherits
this zero-crossing.

\begin{prop}
Assume that $g_0(t)$ satisfies $g_0(\tau)=0$ for
some $\tau \in \mathbb{R}_{0+}$.
Then, $g(\tau)=0$ for any $g \in \mathcal{H}_{G_0}$.
\label{prop:zerocross}
\end{prop}

This proposition
suggests that,
if one knows that the true impulse response
is zero at some time instant $\tau$,
then one should design $G_0(s)$ such that $g_0(\tau)=0$.
A typical case is $\tau=0$, i.e., the relative degree of the
system is known to be higher than or equal to two. For this case, the result can be further strengthened and is shown in the following theorem. 

\begin{thm}
{Assume that the identification input
$u(t)$ is $k-1$ times differentiable, and satisfies }
\begin{align}
\left|\frac{d^i u}{dt^i}\right| < \infty, \quad i=0,1,\ldots,k-1.
\end{align}
If $g_0(t)$ satisfies
$\lim_{t\to +0}\frac{d^j}{dt^j}g_0(t)=0
$ for $j=0, 1, \ldots, k$, then {
$
\lim_{t\to +0}\frac{d^j}{dt^j}{\hat{g}}(t)=0
$ for  $j=0, 1, \ldots, k$.  
}
\label{thm:relativeDegree}
\end{thm}

Moreover, for any $g\in\mathcal H_{G_0}$, how fast $g(t)$ converges to 0 also depends on $g_0(t)$, which is stated in the following theorem.

\begin{thm} \label{thm:convergencerate}
Assume that the input $u(t)$ is bounded, and
let $U \in \mathbb{R}^N$ be a vector whose
$i$-th element is
$\int_0^{t_i}u(t_i-\tau)d\tau$.
When $G_0(s)$ is stable and $G_0(s) \neq 0$,
{$\frac{\hat{g}(t)}{|g_0(t)|}$} converges to
$U^{\top}c$ when $t \to \infty$, where
$c$ is defined as (\ref{eq:c}).
\end{thm}


In summary, $\mathcal{H}_{G_0}$ inherits some properties
of $g_0$, i.e., how $g_0$ crosses or converges to zero.
This is because the linear spline kernel in (\ref{eq:G}) is employed. More specifically,
let $\mathcal{H}_S$ be the RKHS associated with the first order spline kernel (\ref{eq:spline-od1}).
Noting that $f(0)=0$ for any $f\in \mathcal{H}_S$ from the reproducing property,
the properties of $H_{G_0}$ given in this section can be derived accordingly.

\section{MAXIMUM ENTROPY PROPERTY} \label{sec:maxEnt}

Interestingly, the kernel (\ref{eq:G}) also inherits 
the maximum entropy property of the linear spline kernel (\ref{eq:spline-od1}).

\begin{thm}
	
For a given $g_0(t)$ with $t_0, \ldots, t_n$ be a sequence from $\mathbb{R}_{0+}\cup \{\infty\}$,	
let $T_0, \ldots, T_n$ be the permutation of $\{t_0, \ldots, t_n\}$ such that
\begin{align}
0{=}|g_0({T_0})| <  |g_0({T_1})| <  \cdots < |g_0({T_n})|, \label{eq:defT_i}
\end{align}
and consider the stochastic process
defined by
\begin{align}
\begin{gathered}
h^o(T_k)=\sum_{j=1}^k w(j) \sqrt{|g_0(T_{j})|-|g_0(T_{j-1})|}, \\
 k=1, \ldots, n, \quad
h(T_0)=0,
\end{gathered} \label{eq:ho}
\end{align}
where $w(k)$ is a white Gaussian noise with unit variance.
Then, $h^o(T_k)$ is a Gaussian process
with zero mean and $K_{G_0}(T_i, T_j)$ as its covariance function.
In addition, let
$h(T)$ be any stochastic process
defined over $\{T_0, \ldots,T_n\}$ with
$h(T_0)=0$.
Then,
the Gaussian process $h^o$ is the solution of the MaxEnt problem
\begin{align*}
\max_{h(\cdot)}&\ H(h(T_0),\ldots , h(T_n))  \\
{\rm subject}\ {\rm to}&\
\mathbb{E}(h(T_i))=0, i=1, \ldots, n, \\
&\ \mathbb{V}(h(T_{i})-h(T_{i-1}))=|g_0(T_{i})|-|g_0(T_{i-1})|.
\end{align*}
where $H(h(T_0),\ldots , h(T_n))$ denotes the
differential entropy of $[h(T_0),\ldots , h(T_n)]^{\top}$\footnote{The
differential entropy of random variable
$x$ is defined by $-\int p(x) \log p(x) dx$, where
the integral is taken over the support of $p(x)$. }.
\label{thm:maxent}
\end{thm}

This Maximum Entropy interpretation
also suggests the special structure of the inverse of the Gram matrix of $K_{G_0}$.
{For a given $g_0(t)$ with $t_0, \ldots, t_n$ be a sequence from $\mathbb{R}_{0+}\cup \{\infty\}$,	
let $T_0, \ldots, T_n$ be a permutation of $t_0, \ldots, t_n$, which satisfies
(\mbox{\ref{eq:defT_i}})}.
Also let ${\bar{{\bm K}}\in \mathbb{R}^{(n+1)\times (n+1)}}$ be a Gram matrix of 
$K_{G_0}$ defined as 
\begin{align}
{\bar{\bm{K}}}=\begin{bmatrix}
K_{G_0}(t_0,t_0) &  \cdots & K_{G_0}(t_0,t_n) \\
\vdots & \ddots &\vdots \\
K_{G_0}(t_n,t_0) & \cdots& K_{G_0}(t_n,t_n)
\end{bmatrix}.
\end{align}
{Since $|g_0(T_0)|$ is assumed to be zero, $\bar{\bm{K}}$ has a row and a column whose all elements are zero.
We define $\bm{K} \in \mathbb{R}^{n \times n}$ as a matrix constructed by removing such a row and column from $\bar{\bm{K}}$. 
Note that $\bm{K}$ is also a Gram matrix of $K_{G_0}$. 
} 
Before showing the structure of $\bm{K}^{-1}$, we first give the following
result.
\begin{thm} \label{thm:determinant}

The determinant of $\bm{K}$ is given as
\begin{align}
\det\left( \bm{K}\right)= |g_0(T_1)|\Pi_{i=1}^{n-1}\left(
|g_0(T_{i+1})|-|g_0(T_i)|\right).
\end{align}
\end{thm}

Theorem~\ref{thm:determinant} gives the condition where the inverse of $\bm{K}$ exists;
$g_0(t_i)\neq 0$ for all $i$ and $|g_0(t_i)|\neq |g_0(t_j)|$ for all
$i \neq j$.

\begin{thm} \label{thm:invK}
{Let $\bar{g}=[|g_0(t_0)|, \ldots, |g_0(t_n)|]^{\top} \in \mathbb{R}^{n+1}$, and also let 
$g \in \mathbb{R}^n$ be a vector which  
removes the element corresponds to $|g_0(T_0)|$ from $\bar{g}$.} 
Let $R \in \mathbb{R}^{n\times n}$ be a
row-permutation matrix such that 
\begin{align}
{Rg=\begin{bmatrix}
|g_0(T_1)| \\
\vdots \\
|g_0(T_n)|
\end{bmatrix}
\triangleq
\begin{bmatrix}
g_1 \\
\vdots \\
g_n
\end{bmatrix}
.} 
\end{align}
Then
the inverse matrix of $\bm{K}$ is given as
\begin{align}
\bm{K}^{-1}=R^{\top}PR,
\end{align}
where $P \in \mathbb{R}^{n \times n}$ is 
the inverse matrix of the Gram matrix of the first-order spline kernel 
\cite{Chen:2016}, 
\begin{align}
\{P\}_{i,j}=
\begin{cases}
\frac{g_2}{g_1(g_2-g_1)} & i=j=1, \\
\frac{g_{i+1}-g_{i-1}}{(g_{i+1}-g_i)(g_i-g_{i-1})}& i=j=2, \ldots, n-1. \\
\frac{1}{g_n-g_{n-1}} & i=j=n, \\
0 & |i-j|>1, \\
-\frac{1}{\max(g_i,g_j)-\min(g_i,g_j)} & {\rm otherwise},
\end{cases}
\end{align}
\end{thm}

Theorem~\ref{thm:invK} gives the explicit form of the
inverse matrix of $\bm{K}$. 
Note that $P$ is a tri-diagonal matrix. 
This theorem indicates that $\bm{K}^{-1}$ has a sparse structure, i.e.,
it has at most three elements in each row (or column).

\begin{example}
For illustration, we consider the case
$g_0(t)=te^{-t}$, or equivalently, $G_0(s)=\frac{1}{(s+1)^2}$, and show that the corresponding $\bm{K}^{-1}$ has a sparse structure.
We set $t_i = 0.1 \times i \ (i=1, \ldots, 40)$,
and computed $\bm{K}^{-1}$ according to Theorem~\ref{thm:invK}.
\begin{figure}[!t]
\begin{tabular}{ll}
\begin{minipage}[b]{0.45\linewidth}
\centering
\includegraphics[width=4.5cm]{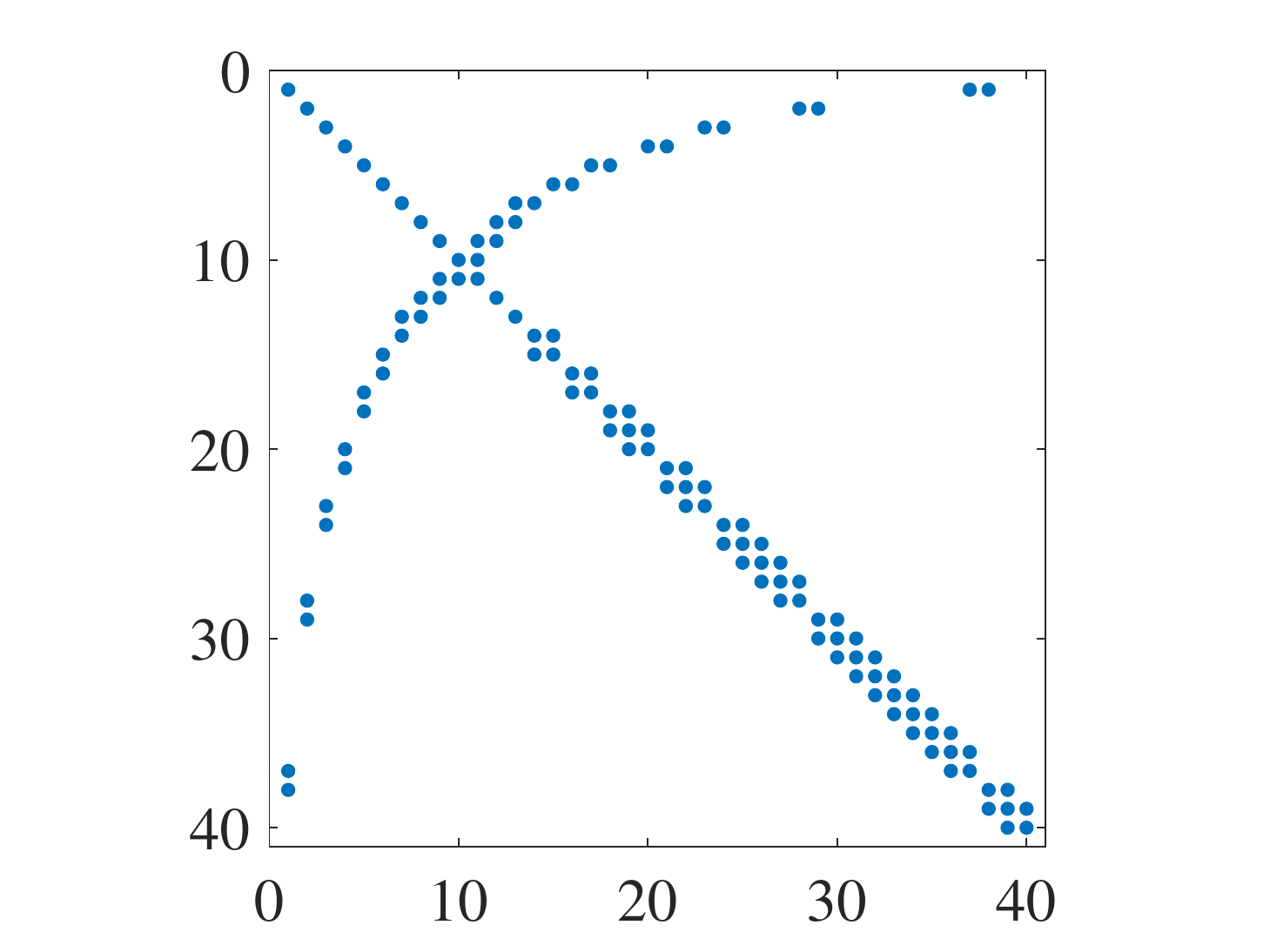}
\subcaption{Sparsity pattern of $\bm{K}^{-1}$}
\label{fig:sparsePattern}
\end{minipage}
\begin{minipage}[b]{0.45\linewidth}
\centering
\includegraphics[width=4.5cm]{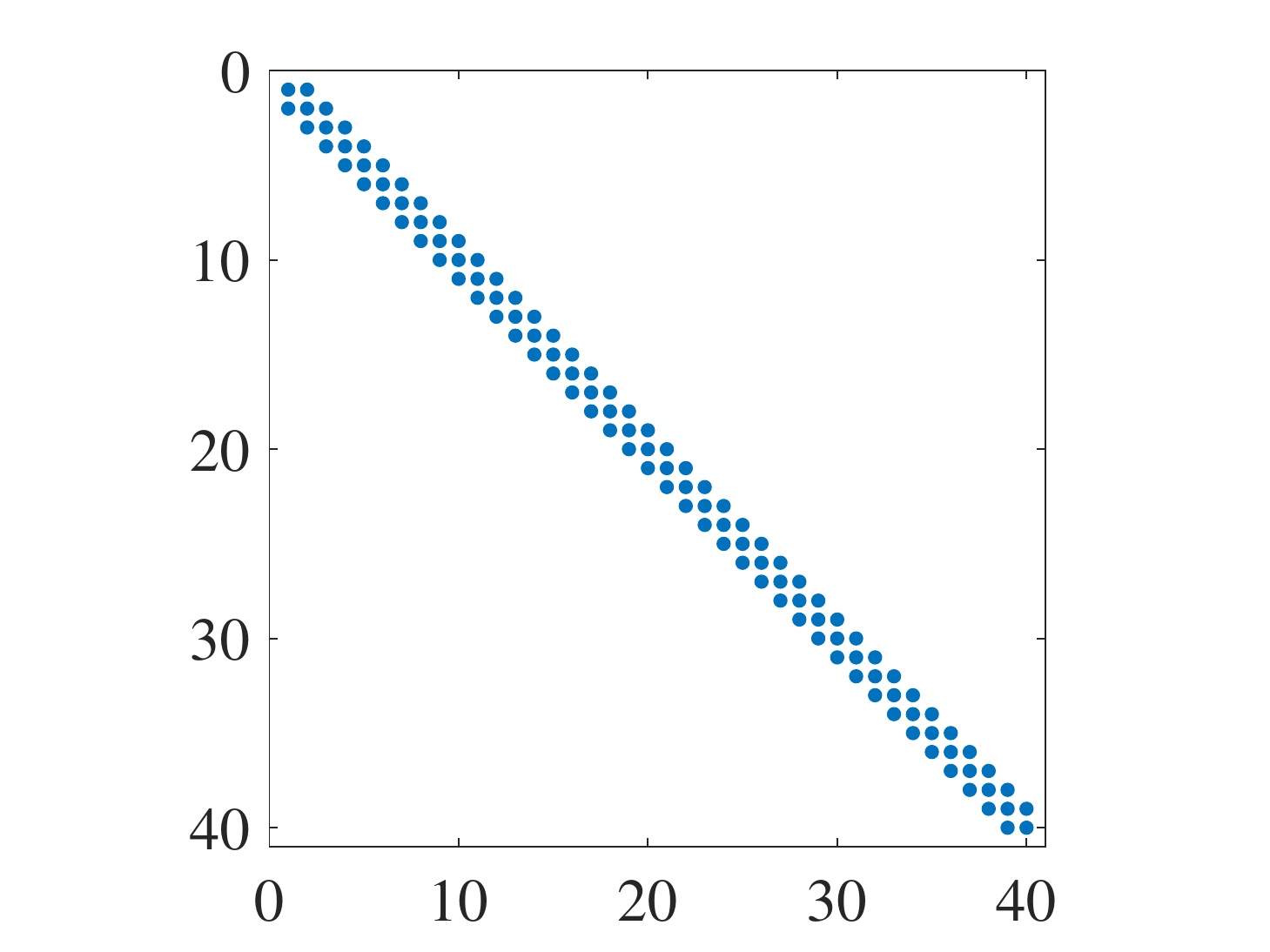}
\subcaption{Sparsity pattern of $P$}
\label{fig:sparsePattern_P}
\end{minipage}
\end{tabular}
\caption{Sparsity pattern of matrices}
\end{figure}
Figs.~\ref{fig:sparsePattern} and \ref{fig:sparsePattern_P} show the sparsity patterns of
$\bm{K}^{-1}$ and $P$, respectively, by using the matlab command {\tt spy}. 
The horizontal and vertical axes show the
column and row of each matrix, respectively,
and the dots show the non-zero elements.
We can see that $P$ is tri-diagonal, and $\bm{K}^{-1}$ has at most three non-zero elements 
in each row or column. 
The sparsity pattern may not be seen in a numerically computed $\bm{K}^{-1}$, e.g., the one computed by using matlab command {\tt inv}.
For instance, {\tt spy(inv(${\bm{K}}$))} shows that all elements
in {\tt inv(${\bm{K}}$)} are non-zero.
To illustrate the effectiveness of Theorem~\ref{thm:invK},
we compute $\left\| 2\times I_{100} - \bm{K} \left(\bm{K}^{-1}\right)'-\left(\bm{K}^{-1}\right)'
\bm{K}\right\|_{\rm FRO}$, where
$\left(\bm{K}^{-1}\right)'$ shows a numerically computed
inverse of $\bm{K}$ with Theorem~\ref{thm:invK} or
{\tt inv}.
Then we have $1.4 \times 10^{-12}$ with Theorem~\ref{thm:invK} and
$1.6 \times 10^{-12}$ with {\tt inv}, respectively.
\end{example}

\section{SPECTRAL ANALYSIS OF MULTIPLE POLE SPLINE KERNEL} \label{sec:SAofMPSK}

It is well-known from Mercer's Theorem that under suitable assumptions on the kernel
any function in the RKHS can be represented by an
orthonormal series.
We show such an orthonormal basis for $\mathcal{H}_{G_0}$, which can yield a reasonable finite dimensional
approximation of $\mathcal{H}_{G_0}$ and can 
make some computations easy and fast.
In this section, we focus on {\mbox{(\ref{eq:G})} where $G_0(s)=\frac{1}{(s+\alpha)^{n+1}}$} with $n=1,2,\ldots$, and show the spectral analysis of (\ref{eq:G}).
This kernel is proposed in \cite{Fujimoto:2017}  and called the Multiple pole Spline kernel.

\subsection{Preliminary}
We first introduce some definitions for a positive
semidefinite kernel
$K: \mathcal{X}\times \mathcal{X}\to \mathbb{R}$ with a compact set $\mathcal{X}$.

Let $\mu$ be a nondegenerate Borel measure on $\mathcal{X}$.
Also let
$L_2(\mathcal{X},\mu)$ denote the space of functions of $f : \mathcal{X} \to \mathbb{R}$
such that
$\int_{\mathcal{X}} |f(x)|^2d\mu(x) < +\infty$.
For a given kernel $K$ and $\phi \in L_2(\mathcal{X},\mu)$, we define an
integral operator on $L_2(\mathcal{X},\mu)$:
\begin{align}
L_K\phi(x) = \int_{\mathcal{X}} K(x,x')\phi(x')d\mu(x'), x\in \mathcal{X}.
\end{align}
If for some $\lambda$,
\begin{align}
L_K \phi(x)=\lambda \phi(x), x \in \mathcal{X},
\end{align}
has the solution other than $\phi(x)=0$,
$\lambda$ and the solution are called the eigenvalues
and eigenfunctions of $L_K$, respectively.
Two distinct eigenfunctions $\phi(x)$ and $\psi(x)$ are orthogonal,
i.e., $\langle \phi , \psi \rangle_{L_2(\mathcal{X},\mu)}=0$.
Then, the kernel $K$ has a series expansion
\begin{align}
K(x,x') = \sum_{i=1}^{\infty} \lambda_i \phi_i(x) \phi_i(x'),\label{eq:KernelExpansion}
\end{align}
which converges uniformly and absolutely on $\mathcal{X}\times \mathcal{X}$.

Consider the first-order spline kernel $K_S(x,x')=\min(x,x'): [0,1]\times [0,1] \to \mathbb{R}$
with $\mu$ being the Lebesgue measure.
In this case, the eigenvalues and eigenfunctions are given by
\begin{align}
&\lambda_i=\frac{1}{\left(i-\frac{1}{2}\right)^2\pi^2}, \phi_i(x)=\sqrt{2}\sin\left(\left(i-\frac{1}{2}\right)\pi x\right), \label{eq:EigenFunctionsForSpline}\\
&\int_0^1 \min(x,x')\phi_i(x')dx'=\lambda_i \phi_i(x)\ (i=1,2,\ldots).
\end{align}
With these $\lambda_i$ and $\phi_i$,
the spline kernel has the series expansion (\ref{eq:KernelExpansion}).

\subsection{Main result} \label{subsec:SpectralAnalysis}
We consider the case $G_0(s)=\frac{\kappa}{(s+\alpha)^{n+1}}$, $n=1,2,\ldots$,
i.e.,
\begin{align}
K_{G_0}(\tau_1,\tau_2)=\min\left(\tau_1^{n}e^{-\alpha \tau_1},\tau_2^{n}e^{-\alpha \tau_2} \right).
\end{align}
For the
simplicity of notations and discussions, we take $\kappa=1$ in the following. 
The extension to other $\kappa\in \mathbb{R}$ is straightforward.
In the rest of this section,
$\lambda_i$ and $\phi_i$ denote the values and functions
defined in (\ref{eq:EigenFunctionsForSpline}).

Before showing the main result,
we first show a lemma.
\begin{lem} \label{lem:EigenFunctionSplineGeneral}
Let $T>0$, and $x \in [0,T]$. Then, $\lambda_i$ and $\phi_i$ defined by (\ref{eq:EigenFunctionsForSpline})
satisfy
\begin{align}
\int_0^T \min(x,x')\phi_i(x'/T)dx'
=T^2\lambda_i\phi_i(x/T).
\end{align}
{In addition, }
\begin{align}
\int_0^T \phi_i(x'/T) \phi_j(x'/T)dx' = 
\begin{cases}
0 & i\neq j, \\
T & i=j.
\end{cases}
\end{align}
\end{lem}
Lemma~\ref{lem:EigenFunctionSplineGeneral}  gives the eigenvalues and eigenfunctions of
$\min(x,x')$ over $(x,x')\in[0,T] \times [0,T]$ for $T>0$. {In particular, 
$\frac{1}{\sqrt{T}}\phi_i(x'/T)$ are orthonormal eigenfunctions. }

The main result of this section is stated as follows.

\begin{thm} \label{thm:spectralAnalysis}
Let $m:\mathbb{R}_{0+}\to \mathbb{R}_{0+}$ be a function defined by
\begin{align}
m(\tau)=\begin{cases}
\frac{1}{2}\tau^{n}e^{-\alpha \tau} & 0\leq \tau \leq \frac{n}{\alpha}\\
\left(\frac{n}{\alpha}\right)^ne^{-n}-\frac{1}{2}\tau^{n}e^{-\alpha \tau}
&\tau \geq \frac{n}{\alpha}
\end{cases},\label{eq:mtau}
\end{align}
and consider the measure induced by $m$; $dm=
\frac{dm}{d\tau}d\tau$ with the Lebesgue measure $d\tau$.
Also let $\lambda_{n,i}$ and $\phi_{n,i}$ be
\begin{align}
\lambda_{n,i}=\left(\frac{n}{\alpha e}\right)^{2n} \lambda_i,
\ \phi_{n,i}(\tau)=\left(\frac{ae}{n}\right)^{\frac{n}{2}}\phi_i\left(\tau^{n}e^{-\alpha \tau} \left(\frac{\alpha e}{n}\right)^{n} \right).
\end{align}
Then, we have
\begin{align}
\int_0^{\infty}\min(\tau_1^n e^{-\alpha \tau_1},
\tau_2^n e^{-\alpha \tau_2})
\phi_{n,i}(\tau_2)dm(\tau_2)
=
 \lambda_{n,i} \phi_{n,i}(\tau_1), \label{eq:EigenFunctionForMSkernel}
\end{align}
with 
\begin{align}
\int_0^{\infty} \phi_{n,i} (\tau) \phi_{n,j}(\tau) dm(\tau)
=\begin{cases}
1 & i=j, \\
0  & i\neq j.
\end{cases}
\end{align}
\end{thm}
Theorem~\ref{thm:spectralAnalysis} suggests that
$\lambda_{n,i}$ and $\phi_{n,i}$ are the eigenvalues and eigenfunctions of
$K_{G_0}$ with the measure induced by $dm$, respectively.

Based on Theorem~\ref{thm:spectralAnalysis}, we have the following theorem. 
\begin{thm} \label{thm:seriesExpansion}
Let $G_0(s)=\frac{1}{(s+\alpha)^{n+1}}$. 
\begin{enumerate}
\item the series expansion 
\begin{align}
K_{G_0}(\tau_1,\tau_2)=\sum_{i=1}^{\infty} 
\lambda_{n,i} \phi_{n,i}(\tau_1) \phi_{n,i}(\tau_2), \label{eq:KG0SeriesExpansion}
\end{align}
converges uniformly and absolutely on $\mathbb{R}_{0+}\times \mathbb{R}_{0+}$. 
\item $\left\{ \sqrt{\lambda_{n,i}} \phi_{n,i}\right\}_{i=1}^{\infty}$ forms an orthonormal basis 
of $\mathcal{H}_{G_0}$, 
and $\mathcal{H}_{G_0}$ has an equivalent representation; 
\begin{align}
\mathcal{H}_{G_0}=
\{
f \mid f(\tau) = \sum_{i=1}^{\infty}
f_i \phi_{n,i}(\tau), 
\sum_{i=1}^{\infty}
\frac{f_i^2}{\lambda_{n,i}}<\infty
\}. 
\end{align}
Moreover, the norm of $f$ is given by 
\begin{align}
\|f\|_{\mathcal{H}_{G_0}}^2
=\sum_{i=1}^{\infty}
\frac{f_i^2}{\lambda_{n,i}}.
\end{align}
\end{enumerate}
\end{thm}

\begin{example}
For illustration, we show the case
with $n=1$ and $\alpha=1$.
\begin{figure}[!t]
\centering
\includegraphics[scale=1]{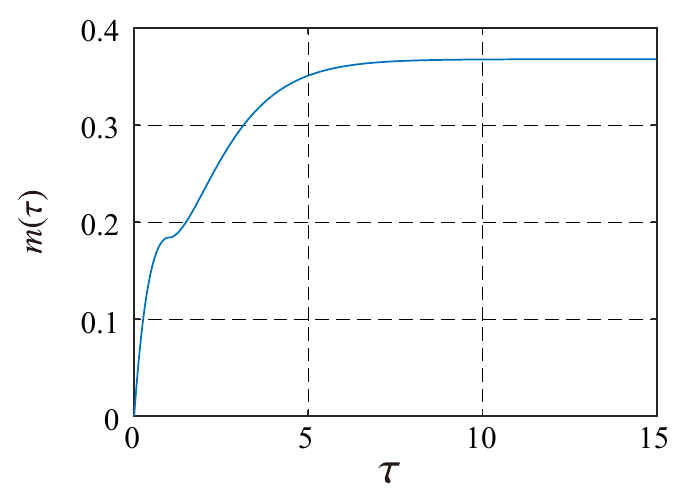}
\caption{Illustration of $m(\tau)$ with $n=1, \alpha=1$}
\label{fig:mt}
\end{figure}
Figs.~\ref{fig:mt} and
\ref{fig:dmdt} shows $m(\tau)$ defined
by (\ref{eq:mtau}) and
$\frac{dm}{d\tau}$, respectively.
The horizontal axes show $\tau$,
and the vertical axes show $m(\tau)$ and
$\frac{dm}{d\tau}$, respectively.
In this case, $\frac{n}{\alpha}=1$
and $\frac{dm}{d\tau}=0$ at
$\tau=1$.
\begin{figure}[!t]
\centering
\includegraphics[scale=1]{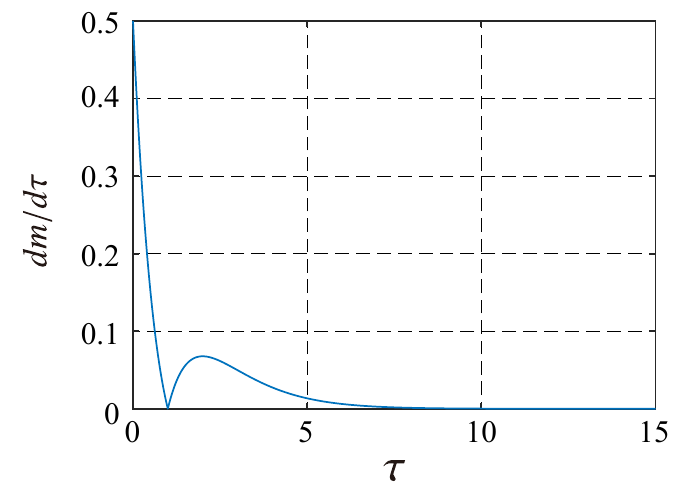}
\caption{Illustration of $\frac{dm}{d\tau}$ with $n=1, \alpha=1$}
\label{fig:dmdt}
\end{figure}
\begin{figure}[!t]
\centering
\includegraphics[scale=1.1]{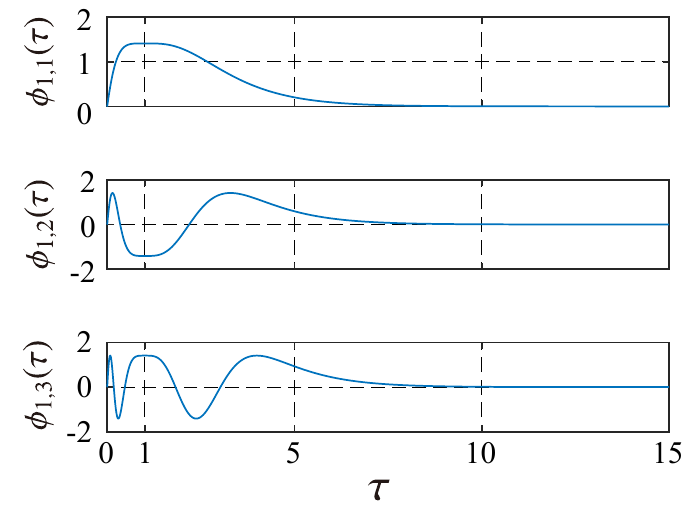}
\caption{Illustration of $\phi_{n,i}(\tau)$ with $n=1,\alpha=1$}
\label{fig:phi_ni}
\end{figure}
Fig.~\ref{fig:phi_ni} shows
$\phi_{1,i}(\tau)$ for $i=1,2,3$.
The horizontal axes show $\tau$, and the vertical axes show
$\phi_{1,i}(\tau)$.
The top, middle, and bottom figures
show $\phi_{1,1}(\tau)$, $\phi_{1,2}(\tau)$,
and $\phi_{1,3}(\tau)$.
These eigenfunctions satisfy
$\phi_{n,i}(0)=0$ and
$\lim_{\tau\to \infty} \phi_{n,i}(\tau)=0$
as we expected. 

{
With the same $n$ and $\alpha$, we also 
compute $\| \bm{K} - \bm{K}_M\|_{\rm FRO}$ where 
$(i,j)$ elements of $\bm{K}$ and 
$\bm{K}_M$ are given as $K_{G_0}(t_i,t_j)$ and 
$\sum_{\ell =1}^M \lambda_{n,\ell } \phi_{n,\ell }(t_i)
\phi_{n,\ell }(t_j)$, respectively, 
with $t_i=0.1 \times i\ (i=1, \ldots, 40)$. 
Fig.~\mbox{\ref{fig:seriesExpansion}} illustrates how $\| \bm{K} - \bm{K}_M\|_{\rm FRO}$ 
converges to zero with increasing $M$. 
The horizontal and vertical axes show $M$ and $\| \bm{K} - \bm{K}_M\|_{\rm FRO}$, 
respectively. }
\begin{figure}[!t]
\centering
\includegraphics[width=8cm]{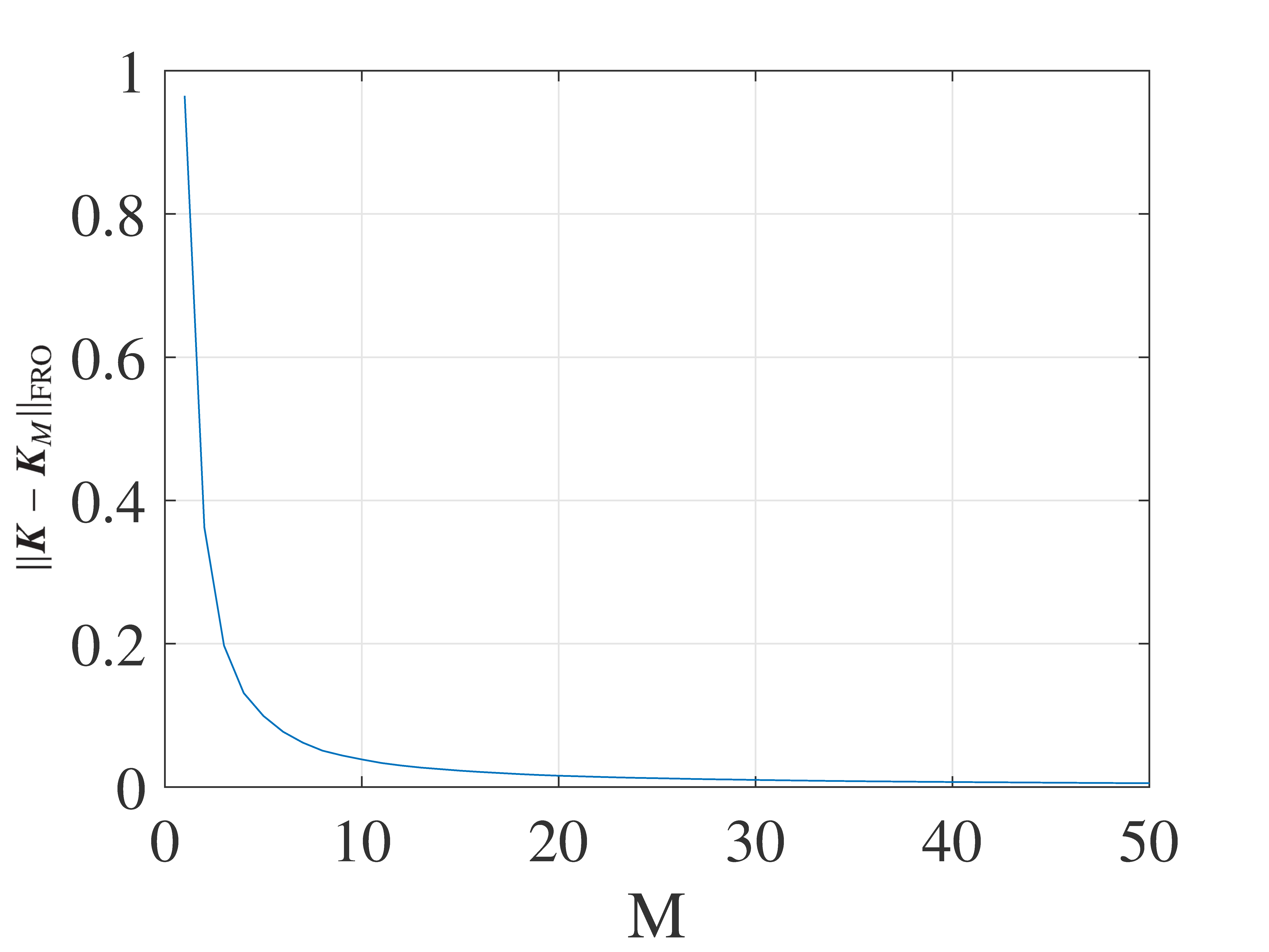}
\caption{Illustration of $\|\bm{K}-\bm{K}_M\|_{\rm FRO}$}
\label{fig:seriesExpansion}
\end{figure}
\end{example}

\section{ILLUSTRATIVE EXAMPLE} \label{sec:sim}
In Sec.~\ref{sec:sim}, we give a numerical example to
illustrate the effectiveness of the proposed kernel.
The target system is
given by
\begin{align}
G^*(s)=\frac{1}{(s+1)(s+3)},
\end{align}
hence the relative degree of the target is two.
For $G_0(s)$,
we employ
\begin{align}
G_0(s)=\frac{\theta_3(\theta_1-\theta_2)}{(s+\theta_1)(s+\theta_2)}
=\theta_3\left(\frac{1}{s+\theta_2}-\frac{1}{s+\theta_1}\right), \label{eq:simG0}
\end{align}
with $\theta=[\theta_1,\theta_2,\theta_3]^{\top} \in \mathbb{R}^3$ as the
hyperparameters of the kernel.
The impulse response of $G_0(s)$ is
\begin{align}
g_0(t)=\theta_3(e^{-\theta_2 t}-e^{-\theta_1 t}),
\end{align}
thus $g_0(0)=0$ for any $\theta$.
As shown in Sec.~\ref{sec:ZeroCross},
this makes the estimated impulse response
$\hat{g}(0)=0$.
This means that we enjoy a priori knowledge
on the system that its relative degree is higher or equal to two.

We consider the case where the input is the impulsive input, and the
noise variance $\sigma^2=10^{-4}$.
The sampling period $T_s$ is set to 0.1 [s], and we
collect $\{y(T_s), y(2T_s), \ldots, y(100T_s)\}=\{y(k T_s)\}_{k=1}^{100}$.
\begin{figure}[!t]
\centering
\includegraphics[scale=1]{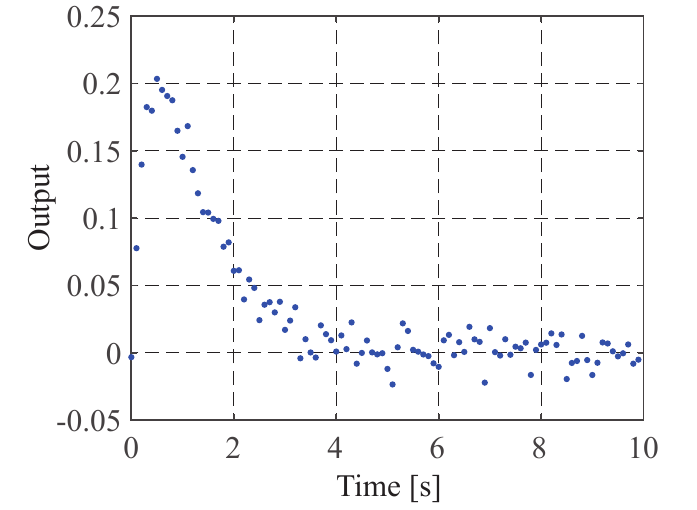}
\caption{Illustration of observed data}
\label{fig:observed}
\end{figure}
Fig.~\ref{fig:observed} shows an example of such observed data $\{y(k T_s)\}_{k=1}^{100}$.
The horizontal axis shows time, and the vertical axis shows the observed output.
Each dot shows the observed data $(i T_s, y(i T_s))$.
In the following, we identify the impulse response from such data
for 300 times with independent noise realizations.

We employ the Empirical Bayes method to tune the hyperparameters,
i.e.,
$\theta$ is tuned so as to maximize
\begin{align}
-\left(\log \det \left(O+\sigma^2 I_N\right)+y^{\top} \left(O+\sigma^2 I_N\right)^{-1}y\right),
\label{eq:marginalLikelihood}
\end{align}
where $\gamma$ is set to $\sigma^2$.
Note that $O$ depends on the hyperparameter $\theta$.
This is based on the Gaussian process interpretation of the kernel based
regularization methods.
In this interpretation, the kernel is regarded as the covariance function of the
zero-mean Gaussian process, and (\ref{eq:marginalLikelihood}) shows the
logarithm of marginal likelihood (some constants are ignored). 
Such a tuning is called the Empirical Bayes \cite{Pillonetto:2014}.

\begin{figure}[!t]
\centering
\includegraphics[width=7.5cm]{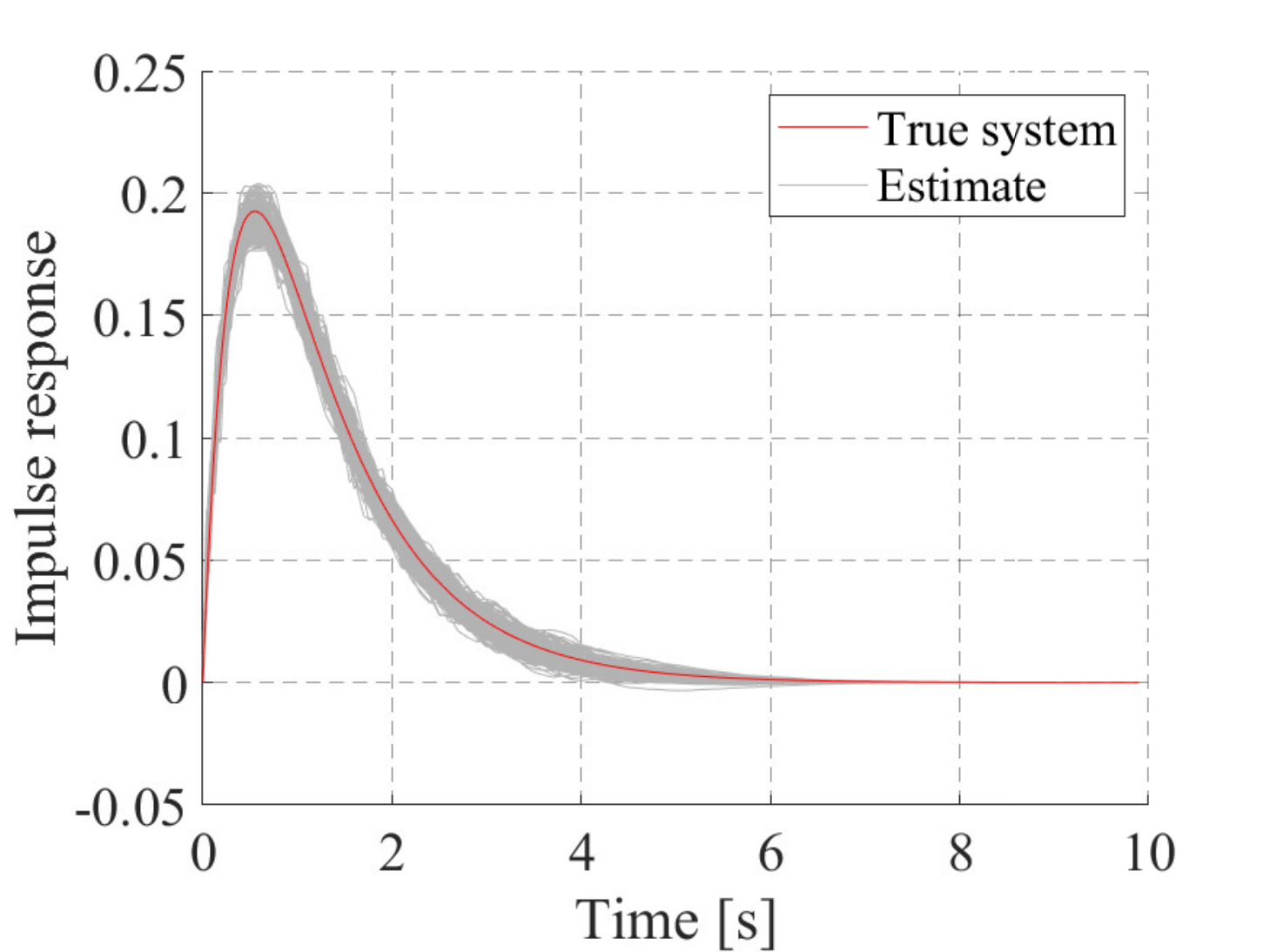}
\caption{Estimated impulse responses with $K_{G_0}$}
\label{fig:proposed}
\end{figure}

Using $G_0(s)$ defined by (\ref{eq:simG0}) and
the Empirical Bayes method, we perform the identification
with $K_{G_0}$ for 300 times with independent noise realizations.
Fig.~\ref{fig:proposed} shows the estimated and true impulse response of the
target system.
The horizontal axis shows time, and the vertical axis shows the impulse response.
The gray lines are 300 estimated impulse responses,
and the red line shows the true impulse response.
Apparently, the behavior of the original impulse response is well
approximated with $K_{G_0}$.

\begin{figure}[!t]
\centering
\includegraphics[width=7.5cm]{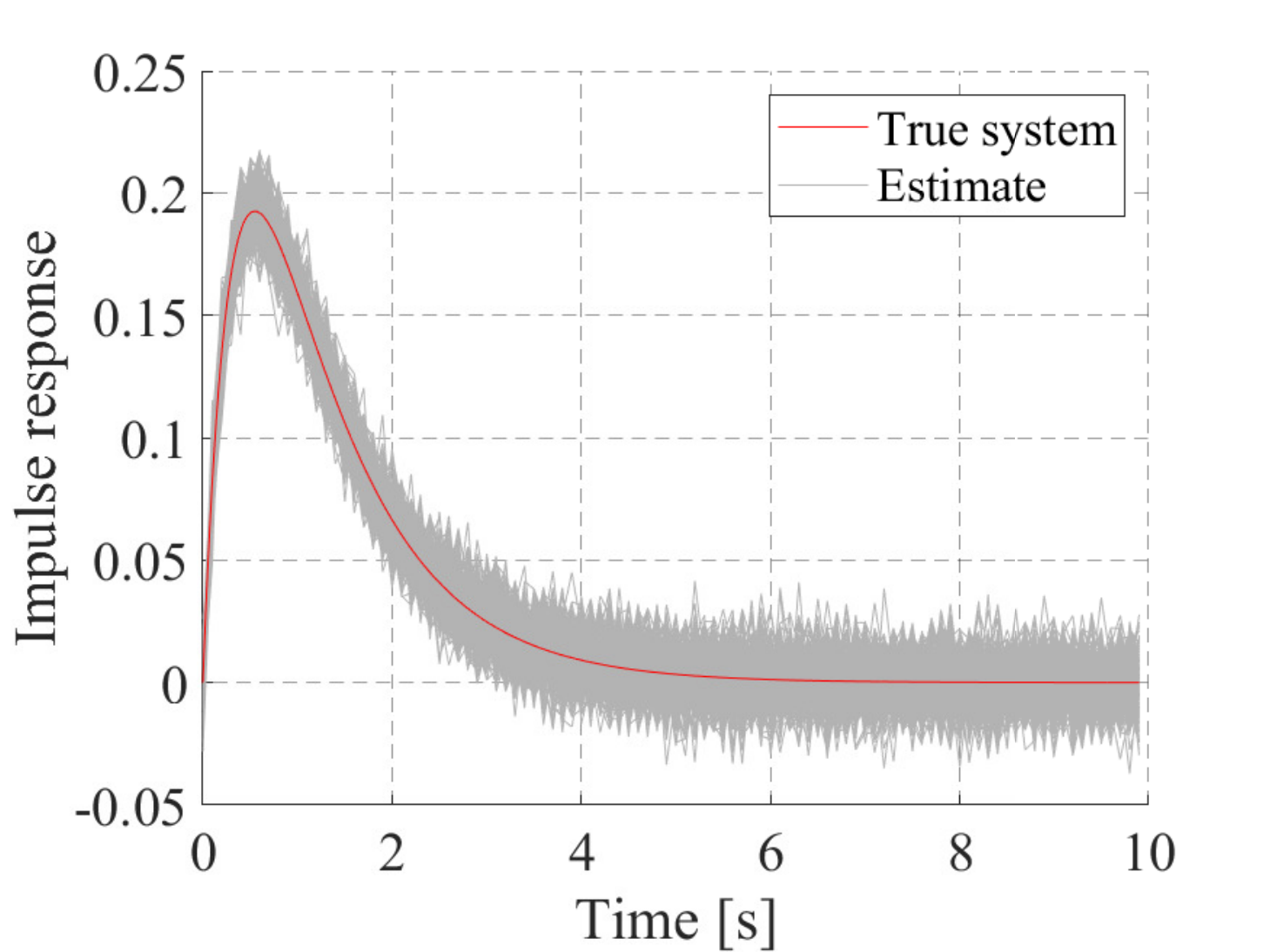}
\caption{Estimated impulse responses with TC kernel (Empirical Bayes)}
\label{fig:EBTC}
\end{figure}

For comparison, we also show the result with
the TC kernel and the Empirical Bayes.
Recall that the TC kernel is defined as (\ref{eq:defTC}). 
Fig.~\ref{fig:EBTC} shows the 100 estimated impulse responses with
the TC kernel and the Empirical Bayes. 
The estimated impulse responses converge to zero slowly, and show overfitting behavior.

For comparison, we also show the results with oracle hyperparameters, i.e., 
hyperparameters tuned with the true impulse response. 
Let $\hat{g}=[\hat{g}(T_s), \hat{g}(2T_s), \ldots, \hat{g}(100 T_s)]^{\top} 
\in \mathbb{R}^{100}$ and 
$g^*=[g^*(T_s), g^*(2 T_s), \ldots, g^*(100 T_s)]^{\top} \in \mathbb{R}^{100}$. 
Noting that we consider the case with impulsive input, we have 
\begin{align}
\hat{g}=&\bm{K}\left(\bm{K}+\sigma^2 I_{100}\right)^{-1}(g^*+w)\nonumber \\
=&\left( I_{100} -\sigma^2\left(\bm{K}+\sigma^2 I_{100}\right)^{-1}\right)(g^*+w),
\end{align}
where $\bm{K}\in \mathbb{R}^{100 \times 100}$ is a Gram matrix of the kernel with $t_i=i T_s$ and 
$w=[w(T_s), w(2T_s), \ldots , w(100T_s)]^{\top} \in \mathbb{R}^{100}$. 
Then, 
\begin{align}
\hat{g}-g^* =& -\sigma^2\left(\bm{K}+\sigma^2 I_{100}\right)^{-1}g^* \nonumber \\
&+\left( I_{100} -\sigma^2\left(\bm{K}+\sigma^2 I_{100}\right)^{-1}\right)w,
\end{align}
and the mean square error on the sampled instants $t_i=iT_s\ (i=1, \ldots, 100)$ 
becomes 
\begin{align}
\mathbb{E}\left[
(\hat{g}-g^*)^{\top}(\hat{g}-g^*)
\right]=&\sigma^4 \left(g^*\right)^{\top}
\left(\bm{K}+\sigma^2 I_{100}\right)^{-2}g^* \nonumber \\
&+100 \sigma^2+ 
\sigma^6 {\rm Tr}\left( 
\left( \bm{K}+\sigma I_{100} \right)^{-2}
\right)\nonumber \\ 
&\quad -2\sigma^4 {\rm Tr}\left( 
\left( \bm{K}+\sigma I_{100} \right)^{-1}
\right). \label{eq:MSE_sampledInstants}
\end{align}
In the following, we show the results with hyperparameters which minimize (\ref{eq:MSE_sampledInstants}). 

\begin{figure}[!t]
\centering
\includegraphics[width=7.5cm]{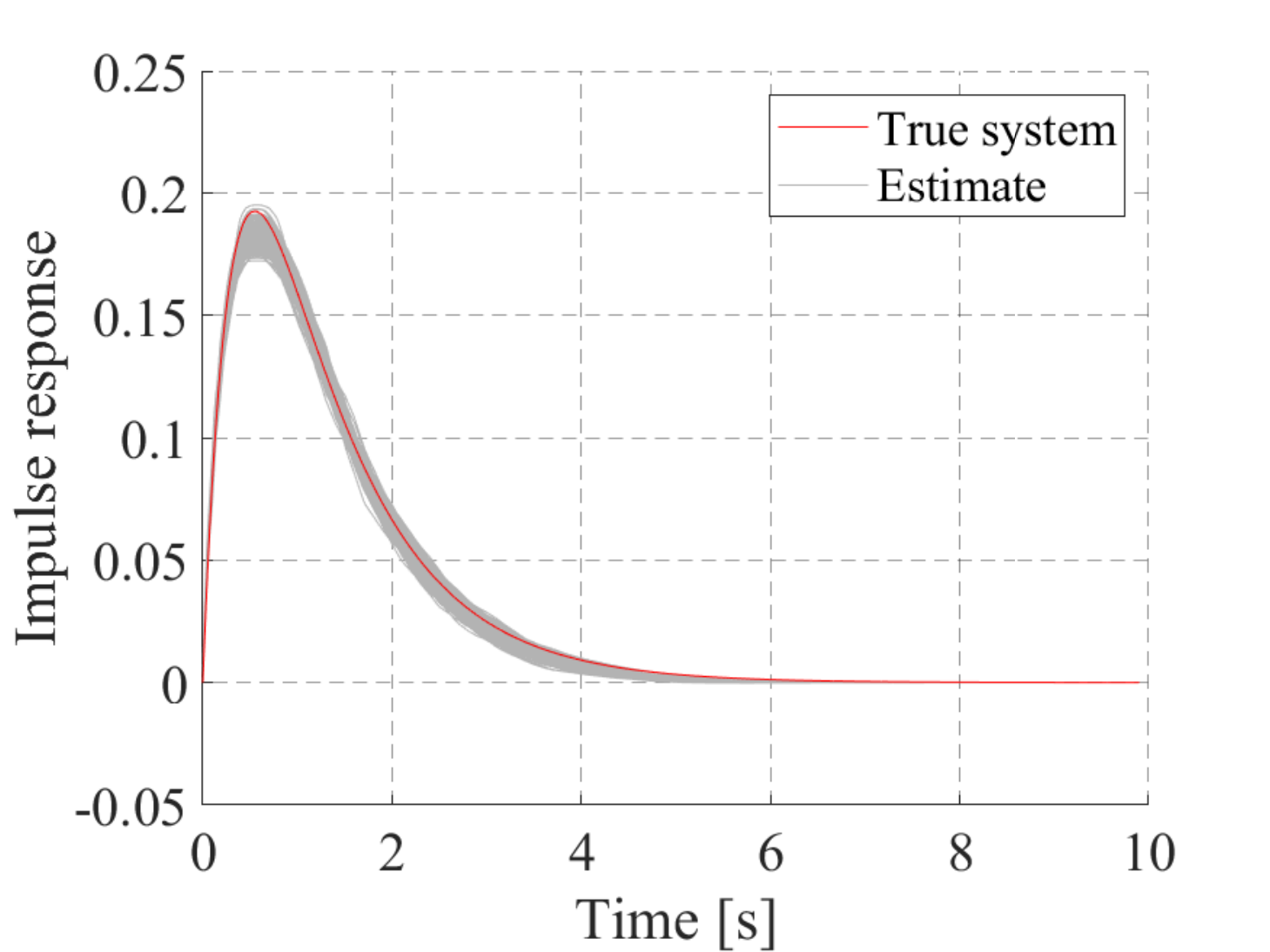}
\caption{Estimated impulse response with $K_{G_0}$ (oracle)}
\label{fig:OracleProposed}
\end{figure}
\begin{figure}[!t]
\centering
\includegraphics[width=7.5cm]{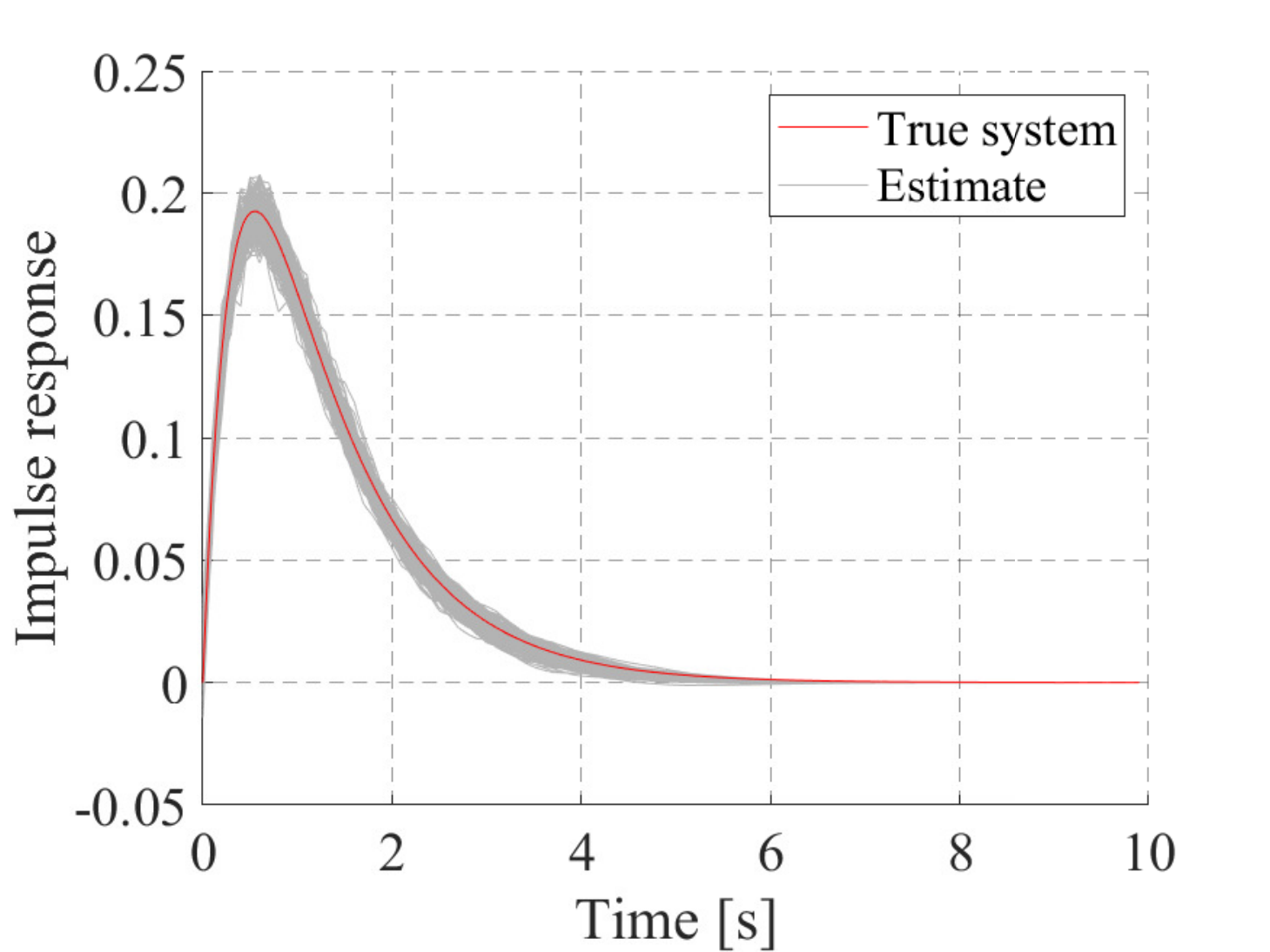}
\caption{Estimated impulse responses with TC kernel (Oracle)}
\label{fig:OracleTC}
\end{figure}

Figs.~\ref{fig:OracleProposed} and \ref{fig:OracleTC} show the 300 estimated impulse responses with such hyperparameters. 
Figs.~\ref{fig:OracleProposed} and \ref{fig:OracleTC} employ the proposed and TC kernel, respectively. 
In this case, the estimated impulse response with the TC kernel converges to zero smoothly. 


\begin{figure}[!t]
\centering
\includegraphics[width=9cm]{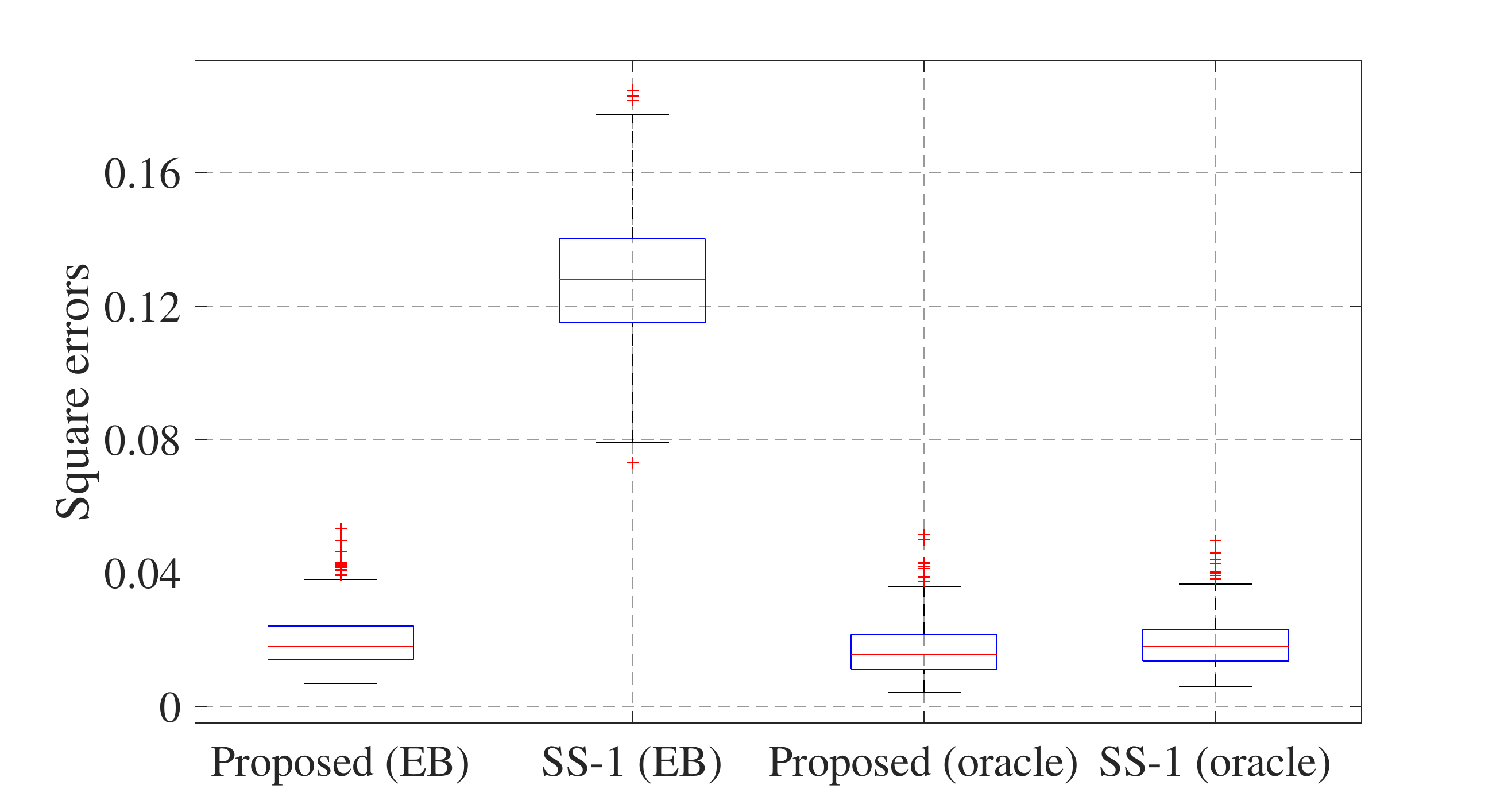}
\caption{Boxplots of square errors on sampled instants}
\label{fig:boxSE}
\end{figure}
Fig.~\ref{fig:boxSE} shows the boxplots of the square errors on the sampled instants, 
i.e., $(\hat{g}-g)^{\top}(\hat{g}-g)$, with 300 independent noise realizations. 
The left two boxes show the results with the Empirical Bayes, and the right two boxes show the results with the hyperparameter tuned according to the mean square error on the sampled instants. 
The proposed kernel with the Empirical Bayes shows almost the same performance as the TC with the oracle hyperparameter, and 
the proposed kernel with the oracle hyperparameter outperforms the others. 
These results show that the proposed kernel is more appropriate for $G^*(s)$ than the TC kernel. 

%

As a statistical analysis, we perform the Wilcoxon rank sum tests for two cases. 
In the first case, we focus on the proposed kernel with the Empirical Bayes and the TC kernel with the oracle hyperparameter. 
The null hypothesis is that two medians of the square errors on the sampled instants are the same (two-sided rank sum test).  
The $p$-value is 0.37, thus this null hypothesis can not be rejected. 
This implies that the proposed method with the Empirical Bayes performs as well as the TC kernel with the optimal hyperparameter. 
In the second case, we focus on the proposed and the TC kernel with the oracle hyperparameters. 
The null hypothesis is that the median of the square errors become smaller with the TC kernel (one-sided rank sum test). 
The $p$-value is $2.0 \times 10^{-4}$, thus the alternative hypothesis is highly significant. 
This suggests that the proposed kernel has potential to achieve better estimate than the TC kernel.


From the above results,
it is confirmed that the prposed kernel (\ref{eq:G}) 
can be useful for regularized impulse resopnse estimation, provided that
the coordinate change is designed by taking into account the a priori knowledge on the system to be identified.

\section{CONCLUSION} \label{sec:conclusion}
This paper focuses on kernels
derived by appling coordinate changes
induced by stable and strictly proper transfer functions to the first-order spline kernel.
They are generalizations of
the tuned-correlated kernel, which is one of the most {widely used} kernels in the
regularized impulse response estimation.
It is shown that the proposed kernels inherit
properties from the coordinate changes
such as the relative degree and the convergence rate.
Also they inherit the Maximum Entropy property from the first-order spline kernel.
Spectral analysis is given for the case where the coordinate change
is chosen as $t^n e^{-\alpha t}$.
Numerical lexample is given to demonstrate the effectiveness of the proposed kernel and shows that a suitable coordinate change could give better performance
than the tuned-correlated kernel.

Extension to cases for the second-order spline kernel or the generalized spline kernel
are future tasks.
Another future task is to find the
optimal coordinate change in some sense for given a priori knowledge on
the system to be identified.

\bibliographystyle{unsrt}

\appendix

\section{Proofs}

\subsection{Proof of Theorem~\ref{thm:pd_stability}}

$K_{G_0}$ {is interpreted as} the first-order spline kernel
with $\beta=\max_t |g_0(t)|$ and
the coordinate change
$\frac{|g_0(t)|}{\max_t |g_0(t)|}: \mathbb{R}_{0+}\to [0,1]$.
This suggests that $K_{G_0}$ is positive definite,
hence there exists an RKHS associated with $K_{G_0}$.

{We recall the following proposition for the
proof about the stability;}
if the kernel $K$ is a nonnegative valued function,
i.e., $K: \mathbb{R}_{0+}\times \mathbb{R}_{0+} \to \mathbb{R}_{0+}$,
then
$K$ is stable if and only if
\begin{align}
\iint_{\mathbb{R}_{0+}^2}
K(\tau_1,\tau_2)d\tau_1d\tau_2<\infty. \label{eq:StableCondition}
\end{align}
See Proposition 15 in \cite{Pillonetto:2014} for more detail about the
stability of the kernel.

The proof about the stability is based on the following Lemma.
\begin{lem} \label{lem:sup_exponential}
For any
stable and strictly proper rational transfer function $G_0(s)$,
there exists $\beta_*>0$ and $\alpha_*>0$
which satisfies
\begin{align}
|g_0(t)|\leq \beta_* e^{-\alpha_* t}\quad \forall t\in \mathbb{R}_{0+}.
\end{align}
\end{lem}
The proof of Lemma~\ref{lem:sup_exponential}
is given in Appendix~\ref{sec:proofLemma}.
Based on Lemma~\ref{lem:sup_exponential},
\begin{align}
\iint_{\mathbb{R}_{0+}^2}
K_{G_0}(\tau_1,\tau_2)d\tau_1 d\tau_2
\leq &
\iint_{\mathbb{R}_{0+}^2}
\beta_*
\min(e^{-\alpha_*\tau_1}
,e^{-\alpha_*\tau_2}
)d\tau_1 d\tau_2
\nonumber \\
=&\frac{2\beta_*}{\alpha_*^2}<\infty.
\end{align}
Since $K_{G_0}$ is a nonnegative valued kernel
and satisfies (\ref{eq:StableCondition}),
the statement is proven.

\subsection{Proof of Lemma~\ref{lem:sup_exponential}} \label{sec:proofLemma}

From the assumption that $G_0(s)$ is stable and
a strictly proper rational function of $s$,
$g_0(t)$ is divided into four parts;
derived from single-real poles,
single-complex poles, repeated real poles,
and repeated complex poles.
In summary, we have
\begin{align}
g_0(t) = &\sum_{i=1}^{N_{\rm real}} \sum_{j=0}^{M_{\rm real}-1} A_{i,j} t^j e^{-\alpha^{\rm real}_ t}
 \nonumber \\ &+\sum_{i=1}^{N_{\rm comp}} \sum_{j=0}^{M_{\rm comp}-1} t^j e^{-\alpha^{\rm comp}_i t}(B_{i,j} \sin \omega_i t +
C_{i,j} \cos \omega_i t), \label{eq:impulse_decompose}
\end{align}
where $N_{\rm real}, N_{\rm comp}, M_{\rm real}$, and $M_{\rm comp}$
denote the number of distinct real poles, the number of distinct complex poles,
the largest multiplicity of the real poles,
and the largest multiplicity of the complex poles, respectively.
$-\alpha^{\rm real}_i \in \mathbb{R}, (i=1,\ldots, N_{\rm real})$ and
and $-\alpha^{\rm comp}_i\pm \omega_i \mathrm{i}, (i=1,\ldots, N_{\rm comp})$ show
the distinct real poles and complex poles, respectively.
Note that $\alpha_i^{\rm real} >0$ and $\alpha_i^{\rm comp}>0$ from the
stability assumption.
In the following, we show that
each term of (\ref{eq:impulse_decompose}) is bounded by an exponential.

For the ease notations, we employ $\alpha$
instead of $\alpha^{\rm real}_i$ for a while.
We show that
$t^{j}e^{-\alpha t} (i\geq 1)$ is bounded by
$j!\left(\frac{2}{\alpha}\right)^{j} e^{-\frac{\alpha}{2}t}$,
where $j!$ denotes the factorial of $j$, i.e.,
$j!=j\times (j-1)\times (j-2)\times \cdots\times2\times1$.
For $\forall t \in \mathbb{R}_{0+}$,
\begin{align}
j!\left(\frac{2}{\alpha}\right)^{j} e^{-\frac{\alpha}{2}t}
-t^j e^{-\alpha t} =&
j!\left(\frac{2}{\alpha}\right)^{j} e^{-\alpha t} \left(e^{\frac{\alpha}{2}t}-
\frac{1}{j!}\left(\frac{\alpha}{2}\right)^{j}t^j \right)
\nonumber \\
=&
j!\left(\frac{2}{\alpha}\right)^{j} e^{-\alpha t} \sum_{k\geq 0, k\neq j}
\frac{1}{k !}\left(\frac{\alpha}{2} t \right)^k
\nonumber \\
\geq &0,
\end{align}
holds.
The second equality is derived from the Taylor expansion of the
exponential function, and
the last inequality is derived from
$\alpha>0, e^{-\alpha t}>0$ and $t\geq 0$.
From this inequality,
we have
\begin{align}
\left|\sum_{i=1}^{N_{\rm real}} \sum_{j=0}^{M_{\rm real}-1} A_{i,j} t^j e^{-\alpha_i^{\rm real} t}\right|
\leq& \sum_{i=1}^{N_{\rm real}} \sum_{j=0}^{M_{\rm real}-1}\left|A_{i,j}\right| \left|t^j e^{-\alpha_i^{\rm real} t}\right| \nonumber \\
\leq& \sum_{i=1}^{N_{\rm real}} \sum_{j=0}^{M_{\rm real}-1}\left|A_{i,j}\right| c_*^{i,j} e^{-\frac{1}{2}\alpha_i^{\rm real} t},
\end{align}
where $c_*^{i,j}=j!\left(\frac{2}{\alpha_i^{\rm real}}\right)^{j}$.
Let $\alpha_*^{\rm real}=\min_i(\frac{1}{2}\alpha_i^{\rm real})$.
Then, $e^{-\frac{1 }{2}\alpha_i^{\rm real}t}\leq e^{-\alpha_*^{\rm real} t}$
for $t\in \mathbb{R}_{0+}$ and
we have
\begin{align}
\left|\sum_{i=1}^{N_{\rm real}} \sum_{j=0}^{M_{\rm real}-1} A_{i,j} t^j e^{-\alpha_i^{\rm real} t}\right|
\leq& \left(\sum_{i=1}^{N_{\rm real}} \sum_{j=0}^{M_{\rm real}-1}\left|A_{i,j}\right| c_*^{i,j}\right) e^{-\alpha_*^{\rm real} t} \nonumber \\
\leq& \beta_*^{\rm real} e^{-\alpha_*^{\rm real}t},
\end{align}
with
\begin{align}
\beta_*^{\rm real}=\sum_{i=1}^{N_{\rm real}} \sum_{j=0}^{M_{\rm real}-1}\left|A_{i,j}\right| c_*^{i,j}.
\end{align}

By noting
\begin{align}
\left|(B_{i,j} \sin \omega_i t +
C_{i,j} \cos \omega_i t)\right|
\leq
\sqrt{B_{i,j}^2+C_{i,j}^2},
\end{align}
the same proof can be applied for the second term of
(\ref{eq:impulse_decompose}), and
\begin{align}
&\left|\sum_{i=1}^{N_{\rm comp}} \sum_{j=0}^{M_{\rm comp}-1} t^j e^{-\alpha^{\rm comp}_i t}(B_{i,j} \sin \omega_i t +
C_{i,j} \cos \omega_i t) \right| \nonumber \\
&\leq \beta_*^{\rm comp} e^{-\alpha_*^{\rm comp} t},
\end{align}
with
\begin{align}
\alpha_*^{\rm comp}=&\min_i \frac{1}{2}\alpha_i^{\rm comp}, \nonumber \\
\beta_*^{\rm comp}=&\sum_{i=1}^{N_{\rm comp}} \sum_{j=0}^{M_{\rm comp}-1}\sqrt{B^2_{i,j}+C_{i,j}^2} j!\left(\frac{2}{\alpha_i^{\rm comp}}\right)^j.
\end{align}

From the above discussions,
we have
\begin{align}
|g_0(t)|\leq&  \beta_*^{\rm real} e^{-\alpha_*^{\rm real} t}+ \beta_*^{\rm comp} e^{-\alpha_*^{\rm comp} t}\nonumber \\
\leq& \beta_* e^{-\alpha_* t},
\end{align}
where
\begin{align}
\beta_*=&2\max \left(\beta_*^{\rm real},\beta_*^{\rm comp}\right), \nonumber \\
\alpha_*=&
\min\left(\alpha_*^{\rm real},\alpha_*^{\rm comp}\right),
\end{align}
and this completes the proof.

\subsection{Proof of Proposition~\ref{prop:zerocross}}
From the reproducing property of $K_{G_0}$,
\begin{align}
g(\tau)=& \InPro{g}{K_{G_0}(\tau,\cdot)} \nonumber \\
=& \InPro{g}{0} =0.
\end{align}
Here we use $K_{G_0}(\tau,t)=\min(0,|g_0(t)|)=0$.

\subsection{Proof of Theorem~\ref{thm:relativeDegree}}

We first prove the case where $k=0$.
Consider $K_{i}^u(t)$ defined by (\ref{eq:ConvKernelInput}).
From the assumption that $g_0(t)\to 0$ when $t\to +0$,
$K_{i}^u(t)$ is rewritten as
\begin{align}
K_{i}^u(t)= \int_0^t u(t_i-\tau) g_0(\tau) d\tau +
\int_t^{t_i} u(t_i-\tau) g_0(t) d\tau, \label{eq:KuiWithg}
\end{align}
for sufficiently small $t$.
By noting $|\int_{0}^{t_i}u(t_i-\tau) d\tau|<\infty$,
we have
$\lim_{t\to +0}K_i^u(t)=0$ from
\begin{align}
\left|K_i^u(t)\right|
\leq
\left|
\int_0^t u(t_i-\tau) g_0(\tau) d\tau
\right| + \left|g_0(t)\right| \left|
\int_t^{t_i} u(t_i-\tau)  d\tau
\right|.
\end{align}
This holds for all $i$, and we
conclude $\lim_{t\to 0}\hat{g}(t) \to 0$.

Next, we consider the case $k=1$.
From (\ref{eq:KuiWithg}), we have
\begin{align}
\frac{d}{dt} K_{i}^u(t) =
\frac{dg_0}{dt} \int_t^{t_i}u(t_i-\tau)d\tau.
\end{align}
Again by noting that $u(t)$ is bounded and
$\frac{dg_0}{dt}\to 0$ from the assumption,
we have
$\lim_{t\to +0} \frac{d}{dt}K_{i}^u=0$ and
$\frac{d\hat{g}}{dt}\to 0$.

Finally, we prove the case where $k\geq 2$.
Let $U_i(t) = \int_t^{t_i}u(t_i-\tau)d\tau$.
When $k\geq 2$,
we have
\begin{align}
\frac{d^k}{dt^k}K_i^u(t)=
\sum_{j=0}^k \frac{d^jg_0}{dt^j}\frac{d^{k-j}U_i}{dt^{k-j}}.
\end{align}
From the assumption that $u(t)$ {and its derivatives are bounded}, the
derivatives of $U_i(t)$ are also bounded {for $j=0, 1, \ldots, k$}.
Thus, if $\frac{d^jg_0}{dt^j}\to0$ for all $j=0, \ldots, k$,
we have
$\lim_{t\to +0}\frac{d^k}{dt^k}K_i^u(t)=0$ and
the proof has been completed.

\subsection{Proof of Theorem~\ref{thm:convergencerate}}
Consider $K_i^u(t)$ defined by (\ref{eq:ConvKernelInput}).
Let
$\mathcal{T}_{1,i} (t) \subset [0,t_i]$ and
$\mathcal{T}_{2,i}(t)  \subset [0,t_i] $ be sets
defined by
$\mathcal{T}_{1,i}(t)=\{\tau \mid
|g_0(t)| \leq |g_0(\tau)|, 0\leq \tau \leq t_i \}$ and
$\mathcal{T}_{2,i}(t)=\{\tau \mid
|g_0(t)| \geq |g_0(\tau)|, 0\leq \tau \leq t_i \}$.
This indicates that
$K_{G_0}(t,\tau)=|g_0(t)|$ when $\tau \in \mathcal{T}_{1,i}(t)$ and
$K_{G_0}(t,\tau)=|g_0(\tau)|$ when $\tau \in \mathcal{T}_{2,i}(t)$.
Hence, we have
\begin{align}
\frac{K_i^u(t)}{|g_0(t)|} = \int_{\mathcal{T}_{1,i}(t)}u(t_i-\tau)d\tau
+\int_{\mathcal{T}_{2,i}(t)}u(t_i-\tau)\left|\frac{g_0(\tau)}{g_0(t)}\right|d\tau.
\end{align}
Note that $0 \leq \left|\frac{g(\tau)}{g(t)}\right|\leq 1$ when $\tau \in
\mathcal{T}_{2,i}(t)$.
Since the integrand of the second term is bounded and
the Lebesgue measure of $\mathcal{T}_{2,i}(t)$ goes to zero
when $t \to \infty$ (because $g_0(t) \to 0$),
\begin{align}
\lim_{t \to \infty} \frac{K_i^u(t)}{|g_0(t)|} = \int_0^{t_i} u(t_i-\tau)d \tau,
\end{align}
and this indicates
\begin{align}
\lim_{t \to \infty} \frac{\hat{g}(t)}{|g_0(t)|}=U^{\top}c.
\end{align}

\subsection{Proof of Theorem~\ref{thm:maxent}}

The former half of the theorem
is easily confirmed by the direct calculation;
\begin{align}
\mathbb{E}[h^o(T_i)h^o(T_j)]=&
\sum_{\ell =1}^{\min(i,j)} |g_0(T_{\ell})|-|g_0(T_{\ell-1})| \nonumber \\
=&|g_0(T_{\min(i,j)})|,
\end{align}
and by noting $|g_0(T_{\min(i,j)})|=\min(|g_0(T_i)|,|g_0(T_j)|)$,
$K_{G_0}$ is the covariance function of $h^o(T_k)$.

The latter half of the theorem is based on
the Lemma~1 of \cite{Chen:2016},
which is stated as follows.
\begin{lem}[Chen \textit{et al}.]
Let $h(t)$ be any stochastic
process with $h(t_0)=0$ for $t_0=0$.
For any $n \in \mathbb{N}$ and
$0=t_0 \leq t_1 \leq \cdots \leq t_n$, the
discrete-time Wiener process is the
solution of the MaxEnt problem
\begin{align*}
\max_{h(\cdot)}&\ H(h(t_0),\ldots , h(t_n))  \\
{\rm subject}\ {\rm to}&\
\mathbb{E}(h(t_i))=0, i=1, \ldots, n, \\
&\ \mathbb{V}(h(t_{i})-h(t_{i-1}))=c(t_i-t_{i-1}),
\end{align*}
where the discrete-time Wiener process is
given by
\begin{align}
\begin{gathered}
h_W(t_0)=0,\ t_0=0, \\
h_W(t_k)=\sum_{i=1}^k w(i-1)\sqrt{t_i-t_{i-1}}, k=1,2,\ldots
\end{gathered} \label{eq:dtWiener}
\end{align}
\end{lem}

Let $g^{\dagger}_0(t)$ be a function which maps
$|g_0(T_i)|$ to $T_i$ for $i=0,\ldots, n$,
i.e., $g_0^{\dagger} (|g_0(T_i)|)=T_i$.
Also let $g_i$ and $h'(g_i)$ be $|g_0(T_i)|$ and $h(g_0^{\dagger}(g_i))=h(T_i)$,
respectively.
With these notations, the original MaxEnt problem becomes
\begin{align*}
\max_{h'(\cdot)}&\ H(h'(g_0),\ldots , h'(g_n))  \\
{\rm subject}\ {\rm to}&\
\mathbb{E}(h'(g_i))=0, i=1, \ldots, n, \\
&\ \mathbb{V}(h'(g_{i+1})-h'(g_i))=g_i-g_{i-1}.
\end{align*}
From Lemma~1 of \cite{Chen:2016},
the optimal solution of this MaxEnt problem
is given by (\ref{eq:dtWiener}), and
this completes the proof.

\subsection{Proof of Theorems~\ref{thm:determinant} and \ref{thm:invK}}

We use the result in \cite{Chen:2016}.
\begin{prop}[Chen \textit{et al.}]
Consider the discrete-time Wiener kernel
\begin{align}
K^{\rm Wiener}(\tau_i, \tau_j) = \min(\tau_i,\tau_j).
\end{align}
Under the assumption that
$0\leq t_1 \leq \cdots \leq t_n < \infty$,
the Gram matrix
\begin{align}
\bm{K}^{\rm Wiener}
=\begin{bmatrix}
K^{\rm Wiener}(t_1,t_1) &  \cdots & K^{\rm Wiener}(t_1,t_n) \\
\vdots & \ddots &\vdots \\
K^{\rm Wiener}(t_n,t_1) & \cdots& K^{\rm Wiener}(t_n,t_n)
\end{bmatrix}.
\end{align}
satisfies
\begin{align}
\det\left( \bm{K}^{\rm Wiener}\right)= t_1 \Pi_{i=1}^{n-1}
(t_{i+1}-t_i),
\end{align}
and
\begin{align}
\left(\bm{K}^{\rm Wiener}\right)^{-1}
=\begin{cases}
\frac{t_2}{t_1(t_2-t_1)} & i=j=1, \\
\frac{t_{i+1}-t_{i-1}}{(t_{i+1}-t_i)(t_i-t_{i-1})}& i=j=2, \ldots, n-1, \\
\frac{1}{t_n-t_{n-1}} & i=j=n, \\
0 & |i-j|>1, \\
-\frac{1}{\max(t_i,t_j)-\min(t_i,t_j)} & {\rm otherwise},
\end{cases}
\end{align}
\end{prop}

By noting that
$R\bm{K}R^{\top}$ is equivalent to
$\bm{K}^{\rm Wiener}$ and $\left(\det(R)\right)^2=1$,
we have the results.


\subsection{Proof of Lemma~\ref{lem:EigenFunctionSplineGeneral}}
With the transformation $X'=x'/T$,
we have
\begin{align}
\int_0^T \min(x,x')\phi_i(x'/T)dx'
=&\int_0^1 \min(x,TX') \phi_i(X')TdX' \nonumber \\
=& T^2\int_0^1  \min(x/T,X') \phi_i(X')dX' \nonumber \\
=& T^2 \lambda_i \phi_i(x/T).
\end{align}

\subsection{Proof of Theorems~\ref{thm:spectralAnalysis} and \ref{thm:seriesExpansion}}

Divide the interval $[0, \infty)$ into
$[0,\frac{n}{\alpha }]$ and $[\frac{n}{\alpha},\infty)$.
Note that $g_0(\tau)=\tau^n e^{-\alpha \tau}$ is monotonic on
each interval from
\begin{align}
\frac{d g_0}{d \tau}=\tau^{n-1}e^{-\alpha \tau}(n-\alpha \tau),
\end{align}
and $g_0(\tau)$ has the inverse function on each interval.
In particular,
the inverse function on $[0,\frac{n}{\alpha }]$
is given by
$Z_p(y) = -\frac{n}{\alpha}W_p(-\frac{\alpha}{n}y^{\frac{1}{n}})$ where
$W_p(x)$ denotes the principal branch of
the Lambert W function (see Appendix~\ref{sec:Wfunction} for a brief introduction of
the Lambert W function).
This is confirmed from the direct calculation;
\begin{align}
g_0(Z_p(y))=&\left(-\frac{n}{\alpha}W_p(-\frac{\alpha}{n}y^{\frac{1}{n}})\right)^n
\exp \left( -\alpha \cdot -\frac{n}{\alpha}W_p(-\frac{\alpha}{n}y^{\frac{1}{n}})\right)\nonumber \\
=&\left(-\frac{n}{\alpha}\right)^n
\left( W_p(-\frac{\alpha}{n}y^{\frac{1}{n}})\right)^n
\left(\exp\left( W_p(-\frac{\alpha}{n}y^{\frac{1}{n}})\right)\right)^n
\nonumber \\
=&\left(-\frac{n}{\alpha}\right)^n \left(-\frac{\alpha}{n} y^{\frac{1}{n}}\right)^n=y,
\end{align}
where $\exp(x)$ denotes $e^x$.
Similarly, the inverse function of $g_0(\tau)$ on the interval
$[\frac{n}{\alpha},\infty)$ is given
by $Z_m(y) = -\frac{n}{\alpha}W_m(-\frac{\alpha}{n}y^{\frac{1}{n}})$ where
$W_m(x)$ denotes the minor branch of
the Lambert W function.
Note that $Z_p(y): [0, \left( \frac{n}{\alpha e}\right)^n]
\to [0, \frac{n}{\alpha}]$ and $Z_m(y): [0, \left( \frac{n}{\alpha e}\right)^n]
\to [\frac{n}{\alpha},\infty)$ satisfy
$m(Z_p(y))=\frac{1}{2}y$ and
$m(Z_m(y))=\left(\frac{n}{\alpha e}\right)^n-\frac{1}{2}y$, respectively.
This indicates $dm(Z_p(y))-dm(Z_m(y))=dy$.

With these inverse relations,
we change the integration variable from $\tau$ to
$y=\tau^n e^{-\alpha \tau}$.
\begin{align}
&\int_0^{\infty}\min(\tau_1^n e^{-\alpha \tau_1},
\tau_2^n e^{-\alpha \tau_2})
\phi_{n,i}(\tau_2)dm(\tau_2) \nonumber\\
=&
 \int_0^{\left(\frac{n}{\alpha e}\right)^n}\min(\tau_1^n e^{-\alpha \tau_1},
y)\left(\frac{ae}{n}\right)^{\frac{n}{2}}
\phi_{i}\left(y\left(\frac{\alpha e}{n}\right)^{n}\right) \left(dm(Z_p(y))-dm(Z_m(y))\right)\nonumber \\
=&
 \int_0^{\left(\frac{n}{\alpha e}\right)^n}\min(\tau_1^n e^{-\alpha \tau_1},
y)\left(\frac{ae}{n}\right)^{\frac{n}{2}}
\phi_{i}\left(y\left(\frac{\alpha e}{n}\right)^{n}\right)dy \nonumber \\
=& \left(\frac{n}{\alpha e}\right)^{2n}  \lambda_i
\left(\frac{ae}{n}\right)^{\frac{n}{2}}\phi_i\left(\tau_1^ne^{-\alpha \tau_1}\left(\frac{\alpha e}{n}\right)^{n} \right)
\nonumber \\
=&\lambda_{n,i}\phi_{n,i}(\tau_1).
\end{align}
Here we use Lemma~\ref{lem:EigenFunctionSplineGeneral}. 
The orthonormality of $\phi_{n,i}(\tau)$ is shown with the same 
integration variable change. 
\begin{align}
&\int_0^{\infty} \phi_{n,i}(\tau) \phi_{n,j}(\tau)dm(\tau) \nonumber \\
=&\left(\frac{ae}{n}\right)^{n}\int_0^{\left(\frac{n}{\alpha e}\right)^n}\phi_i\left(y\left(\frac{\alpha e}{n}\right)^{n}\right)
\phi_{j}\left(y\left(\frac{\alpha e}{n}\right)^{n}\right)dy \nonumber \\
=&\left(\frac{ae}{n}\right)^{n}\int_0^{1}\phi_i\left(y'\right)
\phi_{j}\left(y'\right)\left(\frac{n}{ae}\right)^{n}dy' \nonumber \\
=&\begin{cases}
1 & i=j, \\
0 & i\neq j. 
\end{cases}
\end{align}
The last equality is based on the orthonormality of 
$\phi_i$ over $[0,1]$. 

Theorem~\ref{thm:seriesExpansion} is 
a direct consequence of Theorem~4 in page 37 of  \cite{Cucker:2001}. 

\section{The Lambert W function} \label{sec:Wfunction}
This appendix gives a brief introduction of the Lambert W function.
See e.g., \cite{Corless:1996} for more detail.

The Lambert W function is a set of functions
which satisfies
\begin{align}
z = W(z)e^{W(z)},
\end{align}
for any $z \in \mathbb{C}$.
If we restrict our attention to the case $z \in \mathbb{R}$,
the Lambert W function is divided into two branches; the principal branch and the minor branch.
\begin{figure}[!t]
\centering
\includegraphics[scale=1]{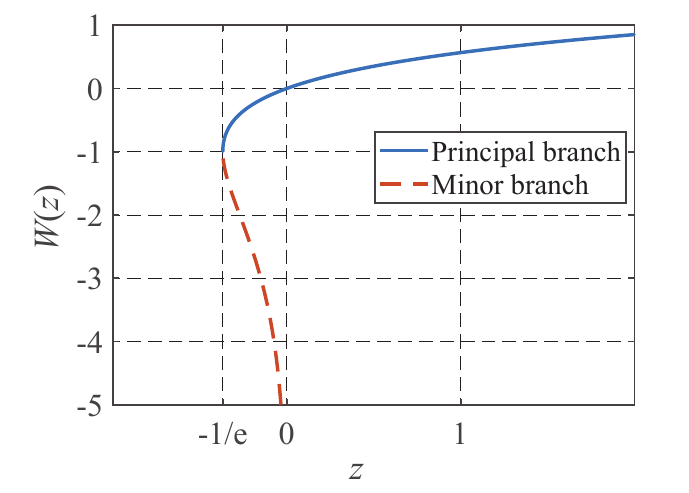}
\caption{Illustration of the Lambert W function}
\label{fig:Wfunction}
\end{figure}
Fig.~\ref{fig:Wfunction} illustrates the Lambert W function on the real axis.
The Lambert W function is double-valued on
$-e^{-1}<z<0$, and divided into two branches;
$W(z)\geq -1$ and $W(z)\leq -1$.
The former one is called the principal branch, and the
latter one is called the minor branch.
We use notations $W_p(z)$ and $W_m(z)$ to denote the
principal and the minor branch, respectively.

\end{document}